\documentclass[twocolumn]{aa}
\usepackage{graphicx}
\usepackage{txfonts}
\newcommand{\srm}{\scriptscriptstyle\rm}
\newcommand{\sit}{\scriptscriptstyle\!\!}
\newcommand{\bgal}{b_{\srm II}}
\newcommand{\lgal}{l_{\srm II}}

\newcommand{\nv}{n_{\srm V}}
\begin{document}

   \title{Mapping extreme-scale alignments of quasar polarization
   vectors\thanks{Based on observations collected at the European
   Southern Observatory 
   (ESO, La Silla and Paranal)}$^{\rm{,}}$\thanks{Table 
   A.1 is only available electronic form at the CDS via anonymous ftp
   to {\tt cdsarc.u-strasbg.fr (130.79.128.5)} or 
   {\tt http://cdsweb.u-strasbg.fr/cgi-bin/qcat?J/A+A/}}}

   \author{D. Hutsem\'ekers\inst{1,}\thanks{Chercheur qualifi\'e
    du F.N.R.S. (Belgique)}, R. Cabanac \inst{2}, H. Lamy \inst{3},
    D. Sluse\inst{1}}

    \institute{Institut d'Astrophysique et de G\'eophysique,
    Universit\'e de Li\`ege, All\'ee du 6 Ao\^ut 17, B5c, B-4000
    Li\`ege, Belgium
    \and Canada France Hawaii Telescope, 65-1238 Mamalahoa Highway,
    Kamuela, Hawaii 96743, USA
    \and BIRA-IASB, Avenue Circulaire 3, B-1180 Bruxelles, Belgium}
  
   \date{Received 29 April 2005; accepted: 27 June 2005}

   \offprints{hutsemekers@astro.ulg.ac.be}   

   \titlerunning{Mapping extreme-scale alignments of quasar polarization 
    vectors} 
   
   \authorrunning{Hutsem\'ekers D. et al.}

\abstract{Based on a new sample of 355 quasars with significant
optical polarization and using complementary statistical methods, we
confirm that quasar polarization vectors are not randomly oriented
over the sky with a probability often in excess of 99.9\%. The
polarization vectors appear coherently oriented or aligned over huge
($\sim$ 1 Gpc) regions of the sky located at both low ($z \sim 0.5$)
and high ($z \sim 1.5$) redshifts and characterized by different
preferred directions of the quasar polarization. In fact, there seems
to exist a regular alternance along the line of sight of regions of
randomly and aligned polarization vectors with a typical comoving
length scale of 1.5 Gpc. Furthermore, the mean polarization angle
$\bar{\theta}$ appears to rotate with redshift at the rate of $\sim$
30$\degr$ per Gpc. The symmetry of the the $\bar{\theta} -z$ relation
is mirror-like, the mean polarization angle rotating clockwise with
increasing redshift in North Galactic hemisphere and counter-clockwise
in the South Galactic one. These characteristics make the alignment
effect difficult to explain in terms of local mechanisms, namely a
contamination by interstellar polarization in our Galaxy. While
interpretations like a global rotation of the Universe can potentially
explain the effect, the properties we observe qualitatively correspond
to the dichroism and birefringence predicted by photon-pseudoscalar
oscillation within a magnetic field. Interestingly, the alignment
effect seems to be prominent along an axis not far from preferred
directions tentatively identified in the Cosmic Microwave Background
maps. Although many questions and more particularly the interpretation
of the effect remain open, alignments of quasar polarization vectors
appear as a promising new way to probe the Universe and its dark
components at extremely large scales.

   \keywords{Quasars: general  -- Polarization -- 
   Large-scale structure of Universe  -- Dark matter}
   }

   \maketitle
%

\section{Introduction}
\label{sec:intro}

Considering a sample of 170 optically polarized quasars with accurate
linear polarization measurements, Hutsem\'ekers ({\cite{HUT98}};
hereafter Paper~I) discovered that quasar polarization vectors are not
randomly oriented over the sky as naturally expected. Indeed, in some
regions of the three-dimensional Universe (i.e. in regions delimited
in right ascension, declination and redshift), the quasar polarization
position angles appear concentrated around preferential directions,
suggesting the existence of very large-scale coherent orientations
--or alignments-- of quasar polarization vectors.

Mainly because the polarization vectors of objects located along the
same line of sight but at different redshifts are not accordingly
aligned, possible instrumental bias and contamination by interstellar
polarization are unlikely to be responsible for the observed effect.
The very large scales at which these coherent orientations are seen
suggest the presence of correlations in objects or fields on spatial
scales up to $\sim$ 1 Gpc at redshift $z \simeq$ 1--2, possibly
unveiling a new effect of cosmological importance.  The interpretation
of such large-scale correlations is difficult within the concordance
cosmological model and might point at missing ingredients. Ongoing
theoretical works (e.g. Das et al. \cite{DAS04}) offer interesting
avenues indicating that we might detect a property of dark matter or
dark energy.

In order to further study the reality of this alignment effect, we
have subsequently carried out a very simple test which consisted in
obtaining new polarimetric measurements for quasars located in a
region of the sky where the range of their polarization position
angles was predicted in advance on the basis of the results of
Paper~I.  These measurements, presented in Hutsem\'ekers \& Lamy
(\cite{HUT01}; hereafter Paper~II), independently confirmed the
existence of coherent orientations of quasar polarization vectors in
the considered region of the sky.  Statistical tests applied to the
whole sample of 213 objects indicate that the quasar polarization
angles are not uniformly distributed with a significance level
(i.e. the probability that the observed statistic is due to chance)
between 10$^{-2}$ and 10$^{-3}$.  These results were confirmed by Jain
et al. (\cite{JAI04}) using coordinate-invariant statistics.

In order to have an accurate and complete description of the
phenomenon, a large number of new polarization measurements is needed.
We have then carried out new polarimetric observations which, combined
to recent data from the literature, lead to a new sample of 355
polarized quasars with accurate linear polarization measurements. In
the following, we present a comprehensive analysis of this new
sample. The characteristics of the sample are described in
Sect.~\ref{sec:sample}. The results of the statistical analysis are
given in Sect.~\ref{sec:stats} and maps of the strongest alignments
are illustrated in Sect.~\ref{sec:maps}.  Possible contamination by
interstellar polarization in our Galaxy is discussed in details in
Sect.~\ref{sec:pism}. The properties of the alignments are
investigated in Sect.~\ref{sec:prop}, and possible interpretations in
Sect.~\ref{sec:inter}.  A preliminary account of this work is reported
in Cabanac et al. (\cite{CAB05}).

\section {The new sample}
\label{sec:sample}

The polarimetric observations were carried out at the European
Southern Observatory (Chile) during five runs in the period August
2000 -- October 2003, using the 3.6m telescope equipped with EFOSC2 in
La Silla and the Very Large Telescope equipped with FORS1 in Paranal.
A detailed account of the observations, data reduction and final
measurements is given in Sluse et al. (\cite{SLU05}), including tests
for possible biases in the data. In total, 184 new, mostly V-band,
polarization measurements were obtained for quasars\footnote{In this
paper we use indifferently the terms ``quasar'' or ``QSO'' for
optically or radio selected quasi-stellar objects.} located at high
galactic latitudes ($|\bgal| \geq 30\degr$) in both the North Galactic
Pole (NGP) and the South Galactic Pole (SGP) regions.  The median
uncertainty of the polarization degree is $\simeq$ 0.25\%.  The
targets were mainly selected from the V\'eron catalogue (V\'eron-Cetty
\& V\'eron \cite{VER01}) and from the Sloan Digital Sky Survey Early
and First Data Releases (Schneider et al. \cite{SCH02,SCH03}, Reichard
et al. \cite{REI03}).  Bright objects were preferred, as well as Broad
Absorption Line (BAL), radio-loud and red quasars which are usually
more polarized (Hutsem\'ekers et al. \cite{HLR98}, Impey \& Tapia
\cite{IMP90}, Smith et al. \cite{SMI02}). Special emphasis has been
given to the observation of quasars located in the direction of the
previously identified regions of polarization vector alignments, i.e.
region A1 located in the NGP region and delimited in (B1950) right
ascensions and redshift by $11^{\rm h}15^{\rm m} \leq \alpha \leq
14^{\rm h}29^{\rm m}$ and $1.0 \leq z \leq 2.3$, and region A3 located
in the SGP region and delimited by $21^{\rm h}20^{\rm m} \leq \alpha
\leq 24^{\rm h}00^{\rm m}$ and $0.7 \leq z \leq 1.5$.  These limits in
right ascension and redshift were fixed visually in Paper~I; it must
be emphasized that they only roughly delineate the true regions of
polarization vector alignments which, ultimately, should be identified
more quantitatively.

The observations were performed with multiple goals in mind: (1) to
reassess the significance of the alignments seen towards the SGP
region A3 as done in Paper~II for the NGP region A1; (2) to increase
the sampling over the high-redshift region A1 where the strongest
alignments are measured and to refine its size by investigating the
alignment in a sub-region; and (3) to increase the sampling in the
foreground regions known to behave differently as shown in Paper~I
and~II.

In the meantime, Smith et al. (\cite{SMI02}) have published new,
mostly unfiltered, polarization measurements for a sample of $\sim$ 70
near-infrared selected QSOs.  These objects are added to our
sample. Most of them are at redshifts $z \leq 0.5$. We also realized
that new redshift measurements were available for a few quasars from
the Impey \& Tapia (\cite{IMP90}) sample used in Paper~I, adding 8
objects to the final sample.

As in Paper~I and~II, we only consider objects which fulfil the
criteria $p \geq 0.6\% $, $\sigma_{\theta} \leq 14\degr$, and $|\bgal|
\geq 30\degr$, where $p$ is the polarization degree and
$\sigma_{\theta}$ the uncertainty of the polarization position angle
$\theta$.  These constraints ensure that most objects are
significantly and intrinsically polarized with little contamination by
the Galaxy, and that the polarization position angles are measured
with a reasonable accuracy (cf. Paper~I for additional details). If an
object has been observed more than once, only the best value is kept
i.e. the measurement with the smallest uncertainty $\sigma_{p}$ on the
polarization degree. Objects flagged as contaminated in Sluse et
al. (\cite{SLU05}) are discarded.

\begin{figure}[t]
\resizebox{\hsize}{!}{\includegraphics*{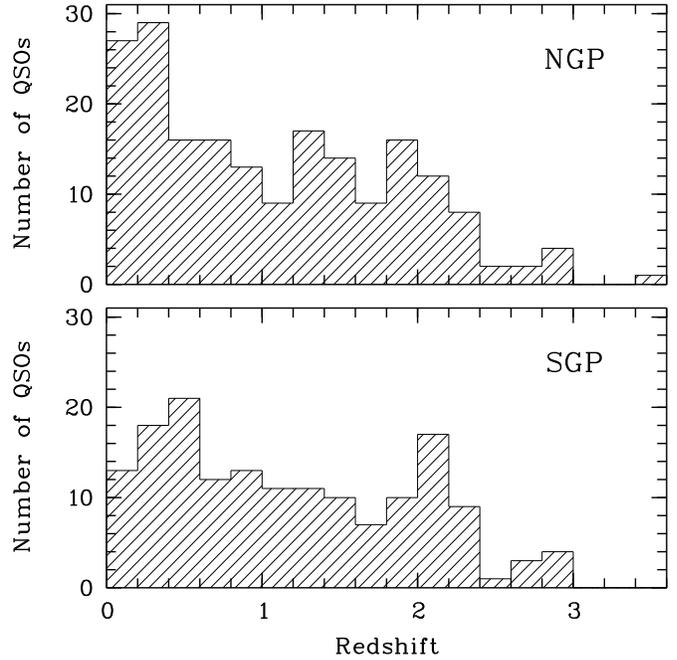}}
\caption{The redshift distribution of the sample of 355 quasars,
illustrated for the NGP and the SGP regions separately.}
\label{fig:hist_z}
\end{figure}

\begin{figure}[t]
\resizebox{\hsize}{!}{\includegraphics*{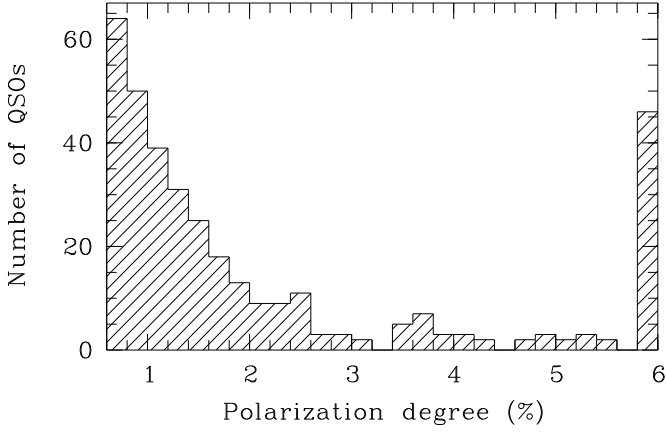}}
\caption{The distribution of the polarization degree for the sample of
355 quasars.  The median is $p \simeq 1.38\%$. The last bin contains
all objects with $p > 6\%$.}
\label{fig:hist_p}
\end{figure}

Combining the new data with the sample of 213 objects from Paper~II,
the final sample of polarized quasars then amounts to 355 objects
distributed all over the sky (195 in the NGP region and 160 in the SGP
region).  The full data set is given in Appendix~\ref{sec:tables}.
The redshift distribution is illustrated in Fig.~\ref{fig:hist_z}; it
shows a good sampling within the range $0 \leq z \leq 2.4$. The
distribution of the polarization degree is illustrated in
Fig.~\ref{fig:hist_p}.

\section {Statistical analysis}
\label{sec:stats}

\subsection{Local statistics}
\label{ssec:statloc}

In this first series of tests we follow the approach of Paper~II, i.e.
we test the hypothesis that the polarization position
angles\footnote{The polarization position angle (or polarization
angle) $\theta$ is expressed in degrees from 0\degr\ to 180\degr\ and
counted from North to East. Polarization vectors (pseudo-vectors in
fact) refer to segments of arbitrary length or normalized on the
polarization degree, centered at the object position and with a
direction fixed by the polarization angle.} of quasars located in a
given region of the sky preferentially lie in the interval [$\theta_1
- \theta_2$] instead of being uniformly distributed. This angular
sector as well as the region of the sky are selected prior to the new
observations, namely on the basis of the results of Paper~I. The
polarization position angles are measured for a sample of quasars
different from that one at the origin of the detection of the effect.
To test the null hypothesis of uniform distribution of circular data
against the alternative of sectoral preference, we use a simple
binomial test (e.g. Lehmacher \& Lienert \cite{LEH80}, Siegel
\cite{SIE56}).  If $P_{\srm A}$ is  the probability under the
null hypothesis that a polarization angle falls in the angular sector
[$\theta_1 - \theta_2$], then $P_{\srm A} = \Delta \theta / 180\degr$
where $\Delta \theta = (\theta_2 - \theta_1) \bmod {180}$.  If $N$
denotes the number of polarization angles falling in [$\theta_1 -
\theta_2$] out of $N_0$ measurements in a given region of the sky, $N$
has a binomial distribution under the null hypothesis and the
probability to have $N_{\star}$ or more polarization position angles
in [$\theta_1 - \theta_2$] is
\begin{equation} 
P_{\rm bin}\,  = \sum_{l=N_{\star}}^{N_0} 
\left(\begin{array}{c}N_0\\l\end{array}\right)
P_{\srm A}^{\,l} \; (1-P_{\srm A})^{N_0 - l}\, . 
\end{equation}

The results of the test are given in Table~\ref{tab:statbin}. For
region A1 we essentially repeat the analysis of Paper~II with
additional data: out of 40 quasars\footnote{Although B1222$+$228 and
B1246$-$047 have recent, better, measurements reported in Table A.1,
these objects were already considered in Paper~I and are then
discarded from the new sample.}  in region A1, 27 have their
polarization angles in the predicted range [146\degr -- 46\degr]
($\Delta \theta = 80\degr$; this range has been defined in
Paper~I). The hypothesis of an uniform distribution of the
polarization position angles is rejected at the 0.3\% level of
significance, which is one order of magnitude smaller than in
Paper~II.  Samples with higher cutoff values of the polarization
degree $p$ are also considered. They show similar departures to
uniformity, indicating that the observed alignments are not only due
to the quasars with the smallest polarization levels.  For
completeness we also provide in Table~\ref{tab:statbin} the numbers
for the full sample, i.e. including the data from Paper~I. In this
case the probability must be seen with caution since the full sample
includes objects at the origin of the detection of the effect.

\begin{table}[t]
\caption{Binomial statistics}
\label{tab:statbin}
\begin{tabular}{lccccc}\hline\hline \\[-0.10in]
Region &  & \multicolumn{2}{c}{New sample} &  
             \multicolumn{2}{c}{Full sample} \\ 
       &  &  $N_{\star}$ / $N_0$  &  $P_{\rm bin}$ 
           & $N_{\star}$ / $N_0$ & $P_{\rm bin}$ \\
       \hline \\[-0.10in]
A1  & $p\geq$ 0.6\% & 27/40 & 2.8 $10^{-3}$ & 42/56 & 3.3 $10^{-6}$ \\
    & $p\geq$ 1.0\% & 15/22 & 2.1 $10^{-2}$ & 27/34 & 3.4 $10^{-5}$ \\
    & $p\geq$ 2.0\% &   5/6 & 6.6 $10^{-2}$ &   7/8 & 1.7 $10^{-2}$ \\
    & $p\geq$ 3.0\% &   5/5 & 1.7 $10^{-2}$ &   6/6 & 7.7 $10^{-3}$ \\[0.1cm]
A1+ & $p\geq$ 0.6\% & 13/14 & 2.2 $10^{-4}$ & 17/18 & 1.1 $10^{-5}$ \\
    & $p\geq$ 1.0\% &   8/8 & 1.5 $10^{-3}$ & 12/12 & 5.9 $10^{-5}$ \\[0.1cm]
A3  & $p\geq$ 0.6\% & 14/18 & 4.3 $10^{-3}$ & 24/29 & 2.6 $10^{-5}$ \\
    & $p\geq$ 1.0\% &  9/11 & 1.3 $10^{-2}$ & 17/20 & 2.3 $10^{-4}$ \\
    & $p\geq$ 2.0\% &   6/7 & 3.3 $10^{-2}$ & 12/13 & 4.6 $10^{-4}$ \\
    & $p\geq$ 3.0\% &   6/7 & 3.3 $10^{-2}$ &  9/10 & 4.1 $10^{-3}$ \\
\hline\\[-0.2cm]\end{tabular}
\end{table}

As pointed out in Paper~I and~II, the polarization vector alignment
seems stronger in the inner part of region A1. We have then defined a
smaller region within A1, denoted A1+, and delimited a priori by
$12^{\rm h}00^{\rm m} \leq \alpha \leq 13^{\rm h}20^{\rm m}$ and $1.3
\leq z \leq 2.0$. Quasars were observed both inside and outside this
region. As seen in Table~\ref{tab:statbin}, nearly all objects located
in region A1+ have their polarization angles in the range [146\degr --
46\degr]. A comparison with the results for the full A1 region
indicates that most of the significance is coming from the inner
region A1+. This supports the fact that this alignment occurs within a
well defined region of the sky. At the same time this illustrates the
difficulty of properly fixing its border.

One of the goal of the new observations was to confirm the
polarization vector alignment in region A3 which is roughly opposite
to A1 on the sky. In Paper~I, we have noted that the polarization
angles of the quasars in region A3 were between 103\degr\ and
144\degr. If we consider a realistic $\Delta \theta = 80\degr$ as for
region A1, we then expect that the polarization angles of quasars
located in region A3 will preferentially fall in the angular sector
[84\degr -- 164\degr]. Out of 18 new polarized quasars in this region,
14 are aligned as expected and the hypothesis of an uniform
distribution of the polarization position angles may be rejected at
the 0.4\% level of significance in favour of coherent
orientation. This confirms the existence of large-scale polarization
vector alignments also for those quasars located in the SGP region A3.

\subsection{Global statistics}
\label{ssec:statglo}

\begin{figure}[t]
\resizebox{\hsize}{!}{\includegraphics*{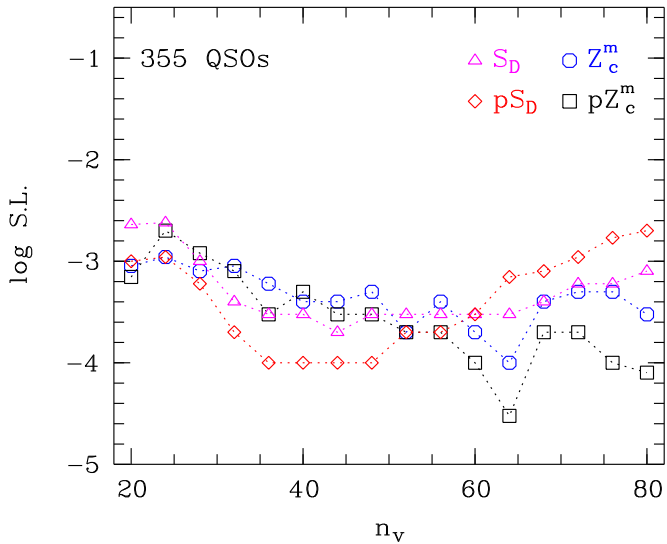}}
\caption{The logarithmic significance level of the statistical tests
applied to the sample of 355 quasars. $\nv$ is the number of nearest
neighbours around each quasar and involved in the calculation of the
statistics.}
\label{fig:sl_355}
\end{figure}

\begin{figure}[t]
\resizebox{\hsize}{!}{\includegraphics*{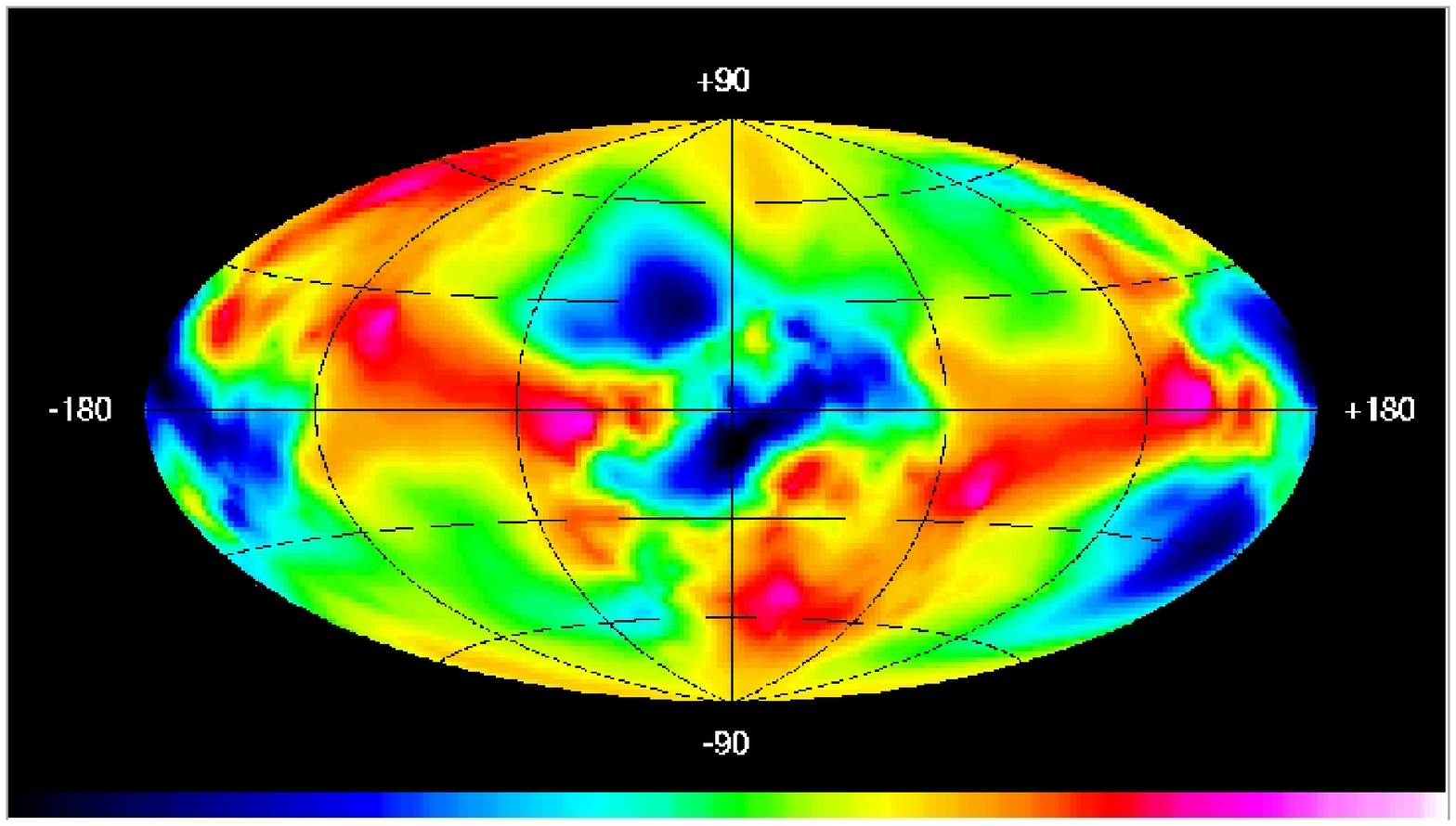}}\\[0.2cm]
\resizebox{\hsize}{!}{\includegraphics*{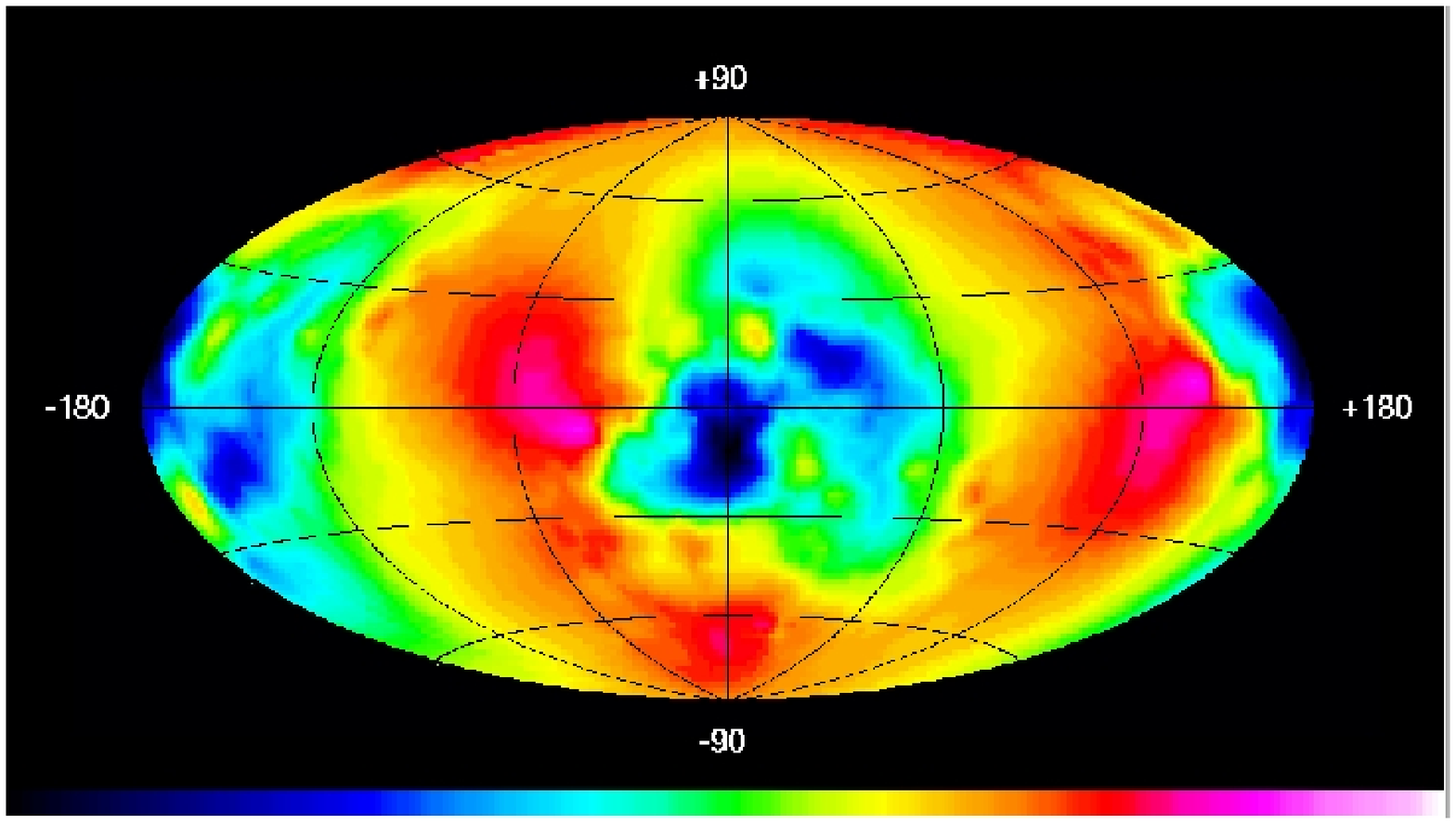}}
\caption{Hammer-Aitoff projection of the $Z_c^m$ (top) and $S_{\sit D}$
(bottom) statistics averaged over $\nv$ = 37 to 43, as a function of
the equatorial coordinates $\alpha_p$ and $\delta_p$ of the northern
pole of an arbitrary coordinate system.  The less significant the
statistics for a given coordinate system or pole position, the darker
the corresponding ($\alpha_p$, $\delta_p$) point in the map.  Note
that ($\alpha_p + 180\degr$, $\delta_p$) is equivalent to ($\alpha_p$,
$- \delta_p$).  The full sample of 355 QSOs is used.}
\label{fig:sl_pole}
\end{figure}

Global statistical tests may be applied to the whole sample to detect
coherent orientations of polarization vectors in some regions of the
sky. Such tests are described in details in Paper~I.  Basically, the
statistics measure the dispersion of the polarization position angles
for groups of $\nv$ nearest neighbours in the 3-dimensional space,
summed over all objects in the sample. The significance is evaluated
through Monte-Carlo simulations, shuffling angles over positions.  A
weakness of the tests used in Paper~I was their dependency upon the
coordinate system. Jain et al. (\cite{JAI04}) made them
coordinate-invariant by incorporating the parallel transport of
polarization vectors.

In the following we consider the $S_{\sit D}$ and the $Z_c^m$ tests
presented in Paper~I. Although it is more sensitive, we do not use
here the $S$ test because it requires an additional parameter. The
tests incorporating parallel transport are denoted $pS_{\sit D}$, and
$pZ_c^m$.  The significance levels (SL) of the statistical tests,
i.e. the probabilities that the test statistics would have been
exceeded by chance only, are computed on the basis of 10$^4$
permutations. When the significance level is smaller than
10$^{-4}$ we used up to 10$^5$ simulations. Significance levels are
given in Fig.~\ref{fig:sl_355} for the new sample of 355 quasars
against the number of nearest neighbours $\nv$ involved in the
calculation of the statistics.

Compared to our previous results (Paper~I and~II), all the statistical
tests indicate a net decrease of the significance level --well below
10$^{-3}$-- for the new, larger, sample (see also Cabanac et
al. \cite{CAB05} for a comparison). This definitely confirms that
quasar polarization vectors are not randomly distributed over the sky
but coherently oriented in groups of $\gtrsim$ 40 objects, i.e. on Gpc
scales at redshift $z\sim 1$. With the increase of the number of
objects, we note a shift of the minimum significance level towards
higher $\nv$.  Tests with and without parallel transport show rather
similar results. This is due to the fact that the groups of quasars
strongly contributing to the significance are located at low
declinations (cf. Sect.~\ref{sec:maps}), i.e. at positions on the
celestial sphere where the corrections for parallel transport remain
small.

As shown in Paper~I, the results of the $S_{\sit D}$ and $Z_c^m$ tests
depend on the adopted coordinate system because the polarization
position angles are defined with respect to meridians.  When projected
onto the equatorial part of the celestial sphere, alignments of
polarization vectors are preserved and well detected by the tests.
On the contrary, if one chooses a coordinate system with a pole
located just in the middle of aligned objects, the polarization angles
will range from 0\degr\ to 180\degr\ and no coherent orientation can
be detected by the tests.  While the parallel transport of
polarization vectors solves this problem, it is nevertheless
interesting to see for which coordinate systems the significance is
extreme.  To investigate this, we have computed the statistics
for various coordinate systems, each one being characterized by a
northern pole of equatorial coordinates $\alpha_p$, $\delta_p$ (see
Paper~I for details and transformation formulae). The results of these
calculations are illustrated in Fig.~\ref{fig:sl_pole}. First, they
confirm that the significance is not extreme in the equatorial
coordinate system ($\delta_p = 90\degr$) and that many systems of
coordinates do provide more significant statistics, a conclusion
already reached in Paper~I.  Interestingly, the statistics show the
lowest significance  when using a coordinate system of northern
pole $\alpha_p \simeq 0\degr$ and $\delta_p \simeq -10\degr$ (which is
equivalent to $\alpha_p \simeq 12^{\rm h}$, $\delta_p \simeq
+10\degr$).  The location of this pole corresponds to the centers of
regions A1 and A3 which are roughly opposite on the sky\footnote{When
projected onto the sky, the center of region A1 is close to the center
of the Local Supercluster (Paper~II). The position ($\alpha_p \simeq
12^{\rm h}$, $\delta_p \simeq +10\degr$) is within a few degrees from
the Virgo cluster, known to be at the center of the Local Supercluster
(e.g. Vall\'ee \cite{VAL02}).}.  Since putting a polar axis at this
location scrambles the most significant alignments, this clearly
suggests that regions A1 and A3 are major contributors to the global
significance.  This is independently verified by considering the
sample with and without the 183 objects along the ``A1--A3 axis'' (as
defined in Sect.~\ref{sec:maps}): while a strong departure to
uniformity is observed when only those quasars belonging to the
A1--A3 region are considered, no significant effect is detected when
these objects are removed from the sample.

\begin{figure}[t]
\resizebox{\hsize}{!}{\includegraphics*{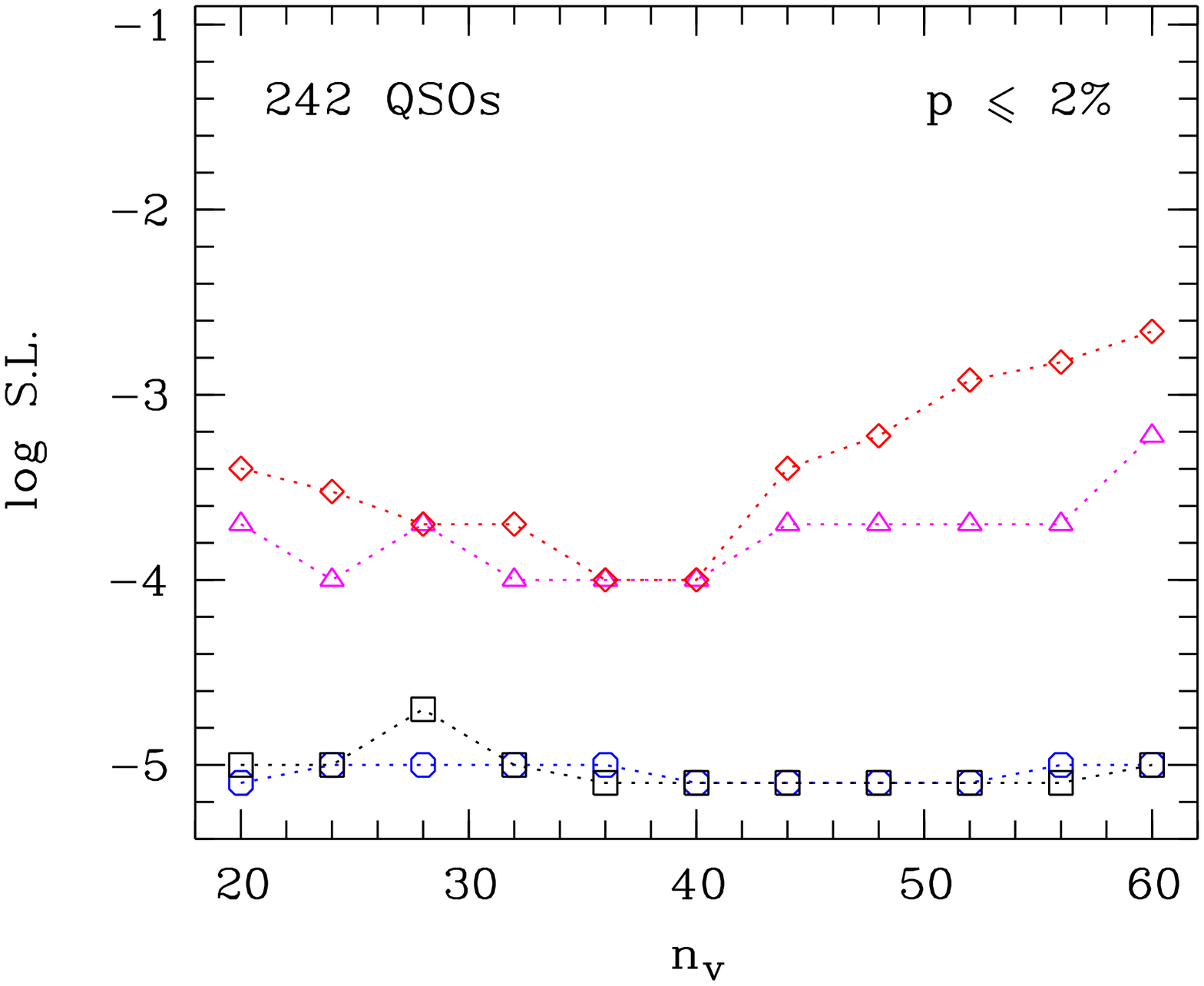}}\\[0.3cm]
\resizebox{\hsize}{!}{\includegraphics*{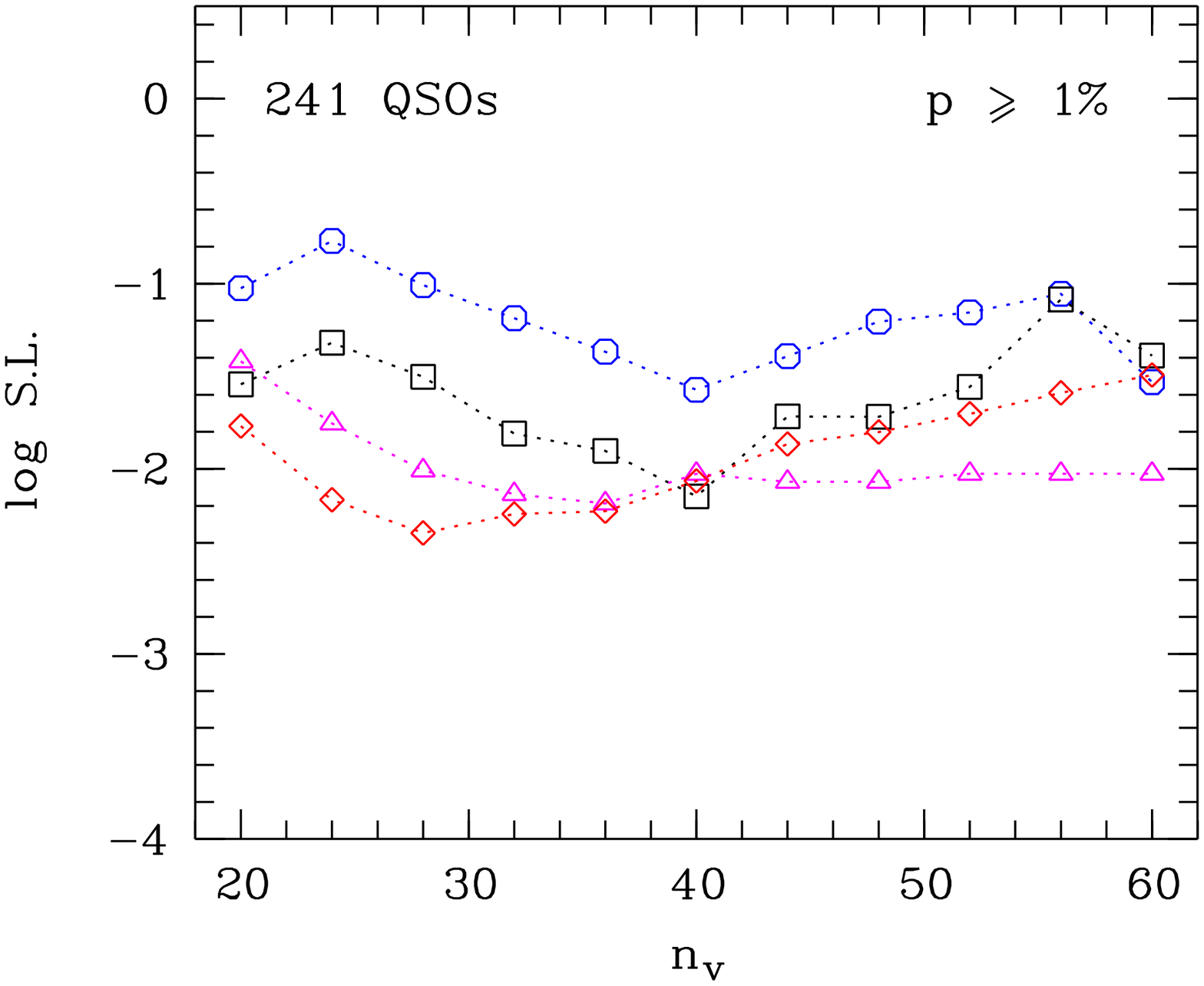}}
\caption{Same as Fig.~\ref{fig:sl_355} but for two sub-samples with
different cutoffs of the polarization degree.  When SL is smaller than
10$^{-5}$ (i.e. unresolved with 10$^{5}$ simulations), we arbitrarily
fix its value to SL = 8~10$^{-6}$.}
\label{fig:sl_p}
\end{figure}

In Fig.~\ref{fig:sl_p}, we give the significance levels of the tests
considering two sub-samples with cuts on the polarization degree. If
only low-polarization ($p \leq 2\%$) quasars are considered, the
departure to uniformity from the $Z_c^m$ and $pZ_c^m$ tests is
stronger than in the full sample. For the $p \geq 1\%$ sub-sample, the
departure to uniformity becomes weakly significant.  These results are
consistent with those obtained by Jain et al. (\cite{JAI04}), although
the differences we note within our new sample are not as strong as
theirs. These differences may indicate that the alignment effect is
more efficient for the low polarization quasars than for the high
polarization ones.  Another reason could be a blurring of the
alignments by the high polarization quasars due to the fact that these
objects are often variable in both polarization degree and angle
(e.g. Impey \& Tapia \cite{IMP90}, a good example being \object{PKS~1216$-$010}
discussed in Sluse et al. \cite{SLU05}).  However this behavior seems
at odds with the results from local statistics
(Table~\ref{tab:statbin}) which indicate that high polarization
objects are aligned as the low polarization ones, namely in regions A1
and A3.  In fact it is important to realize that cutting at $p \geq 1
\%$ removes proportionally more objects located in the regions of
strong alignment A1 and A3 than outside these regions, which results
in a decrease of the global significance. Inversely, cutting at $p
\leq 2 \%$ removes proportionally less objects within these regions,
then increasing the global significance. The fact that more
low-polarization objects lie in regions A1 and A3 is partly due to the
way we have selected the objects. For example, when we got additional
data to confirm the alignment in region A1, we preferentially observed
BAL QSOs whose polarization levels peak close to 1\% (Hutsem\'ekers \&
Lamy \cite{HUT02}, Schmidt \& Hines \cite{SCH99}).  Due to such
intricate selection effects, the results of the tests applied to
sub-samples must be seen with caution.

The same kind of bias occurs when we cut on redshift.  Jain et
al. (\cite{JAI04}) have divided the sample of 213 objects between $z
\geq 1$ and $z \leq 1.3$. With $pS_{\sit D}$-type tests, they have
noted a stronger alignment effect in the high redshift sub-sample, and
no alignment at all in the low redshift one.  In fact, when building
the sample of 213 objects presented in Paper~II, we have mainly added
high-redshift objects in region A1, while in the current paper we also
add many objects at lower redshifts.  As a consequence, when
cutting in redshift the new sample of 355 objects, the differences of
significance between the low and high redshift sub-samples are not as
strong as those reported by Jain et al. (\cite{JAI04}).  Namely, the
tests applied to the new data do not indicate a much higher
significance in the high-redshift sample, and a clear signal is seen
in the low-redshift one.  Cuts on redshift are further discussed in
the next section.

\subsection{Semi-global statistics}
\label{ssec:statsem}

\begin{figure}[t]
\resizebox{\hsize}{!}{\includegraphics*{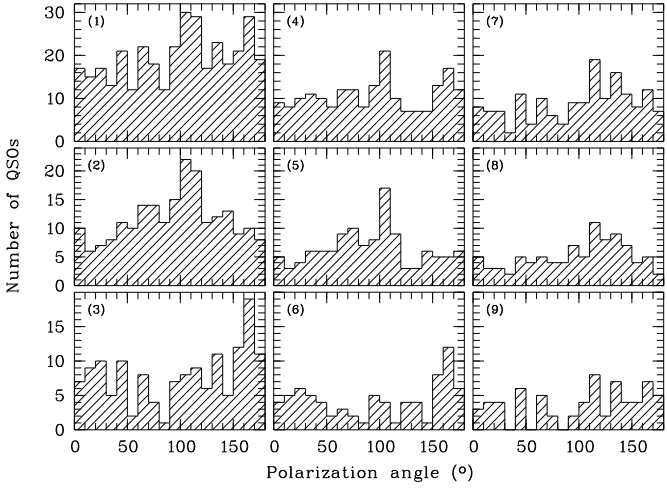}}
\caption{Distributions of the quasar polarization position angles for
different sub-samples. The labels refer to the samples defined in
Table~\ref{tab:statsemi}.}
\label{fig:hist_s}
\end{figure}

Within the sample of 355 quasars, the polarization angles do not
appear uniformly distributed, namely when applying a cut on redshift
as shown in Fig.~\ref{fig:hist_s} (see also Figs.~\ref{fig:mapn}
\&~\ref{fig:maps}) . The isotropy of the histograms are analysed using
the Hawley-Peebles statistical test which also provides an estimate
of the preferred orientations.

The Hawley-Peebles Fourier method (Hawley \& Peebles \cite{HAW75}; see
also Paper~I) is based on fitting the observed distribution by a model
of the form $N(\theta_i) = \overline{N}\, (1+\Delta_1 \cos 2 \theta_i
+ \Delta_2 \sin 2 \theta_i)$ where $\overline{N}$ is the mean number
of objects per bin (we adopt 18 bins); $\Delta_1$ and $\Delta_2$
denote the coefficients of the wave model which describe the degree of
deviation from isotropy.  The probability that the total amplitude
$\Delta = (\Delta_1^2 + \Delta_2^2)^{1/2}$ exceeds some chosen value
is computed to be $P_{\srm \! HP}$ = $\exp \, (-0.25\, n \, \Delta^2$)
where $n$ is the number of objects in the sample. The preferred
orientation is calculated from $\overline{\theta} = 0.5 \arctan \,
(\Delta_2/\Delta_1)$.  Results are given in Table~\ref{tab:statsemi}
for the sub-samples illustrated in Fig.~\ref{fig:hist_s}. We also
used the Rayleigh test (e.g. Fisher \cite{FIS93}) which is very
similar and gives nearly identical results.

The most remarkable result from Fig.~\ref{fig:hist_s} and
Table~\ref{tab:statsemi} is that both the low and high redshift quasar
samples show non-uniform distributions of their polarization angles
and different preferred directions.  The weak anisotropy in the
all-$z$ sample reflects the relative proportion of the various
sub-samples. In the NGP region ($\bgal \geq +30\degr$), the all-$z$
sample is essentially randomly oriented while the low- and high-$z$
samples have very different distributions.  The situation is less
clear for the SGP region ($\bgal \leq -30\degr$): while a definite
anisotropy is seen in the low-$z$ sample, the evidence for a different
preferred direction in the high-$z$ sample is weak, possibly due to
the smaller sample.  It should be emphasized that this behavior does
not mean that all quasars at high or low redshifts have their
polarization vectors coherently aligned.  Indeed, the observed
anisotropy is mainly due to the objects located in the regions of
alignment which have been preferentially targetted, as verified by
running the test after removing these objects.

\subsection{Statistical tests: summary}
\label{ssec:statcon}

The new sample of 355 quasars has been analysed using various,
complementary, statistical methods. All of them concur to indicate
that quasars polarization vectors are definitely not randomly oriented
but coherently oriented over very large spatial scales. With respect
to previous work, the probability that the alignments are due to
chance is definitely lower, often smaller than 10$^{-3}$.

Towards the NGP, we confirm the high significance of the alignment
seen in the high-redshift region A1. A significant alignment --with a
different preferred direction-- is also detected at lower redshift
(see also Fig.~\ref{fig:mapn}).  In the SGP region, we confirm the
alignment previously suspected in region A3. These regions appear as
major contributors to the global significance.

Due to its restricted extension in redshift and heterogeneous density,
the present sampling does not allow us to study the statistical
properties of alignment structures over large volumes.  Yet we see
clear trends for alignments with different preferred directions to
occur in well-defined, although loosely delimited, regions of the sky.
Further characterization will require a denser and larger sampling.

\begin{table}[t]
\caption{Results of the Hawley-Peebles test}
\label{tab:statsemi}
\begin{tabular}{lrcrcr}\hline\hline \\[-0.10in]
Sample & $n$ & $P_{\srm \! HP}$ & $\overline{\theta}$ (\degr) \\
\hline \\[-0.10in]
(1) \ All                                    & 355 & 2.6 10$^{-2}$ & 128 \\
(2) \ $z \leq 1.3$                           & 211 & 2.9 10$^{-4}$ & 104 \\
(3) \ $z \geq 1.3$                           & 144 & 8.9 10$^{-4}$ & 165 \\
(4) \ $\bgal \geq +30\degr$                  & 195 & 8.2 10$^{-1}$ & 107 \\
(5) \ $\bgal \geq +30\degr$, $z \leq 1.3$    & 118 & 4.7 10$^{-3}$ &  90 \\
(6) \ $\bgal \geq +30\degr$, $z \geq 1.3$    &  77 & 4.9 10$^{-3}$ & 175 \\
(7) \ $\bgal \leq -30\degr$                  & 160 & 1.9 10$^{-3}$ & 132 \\
(8) \ $\bgal \leq -30\degr$, $z \leq 1.3$   &  93 & 4.2 10$^{-3}$ & 119 \\
(9) \ $\bgal \leq -30\degr$, $z \geq 1.3$   &  67 & 4.6 10$^{-2}$ & 151 \\
\hline\\[-0.2cm]\end{tabular}
\end{table}

\section {Maps of the alignments}
\label{sec:maps}

\begin{figure}[t]
\resizebox{\hsize}{!}{\includegraphics*{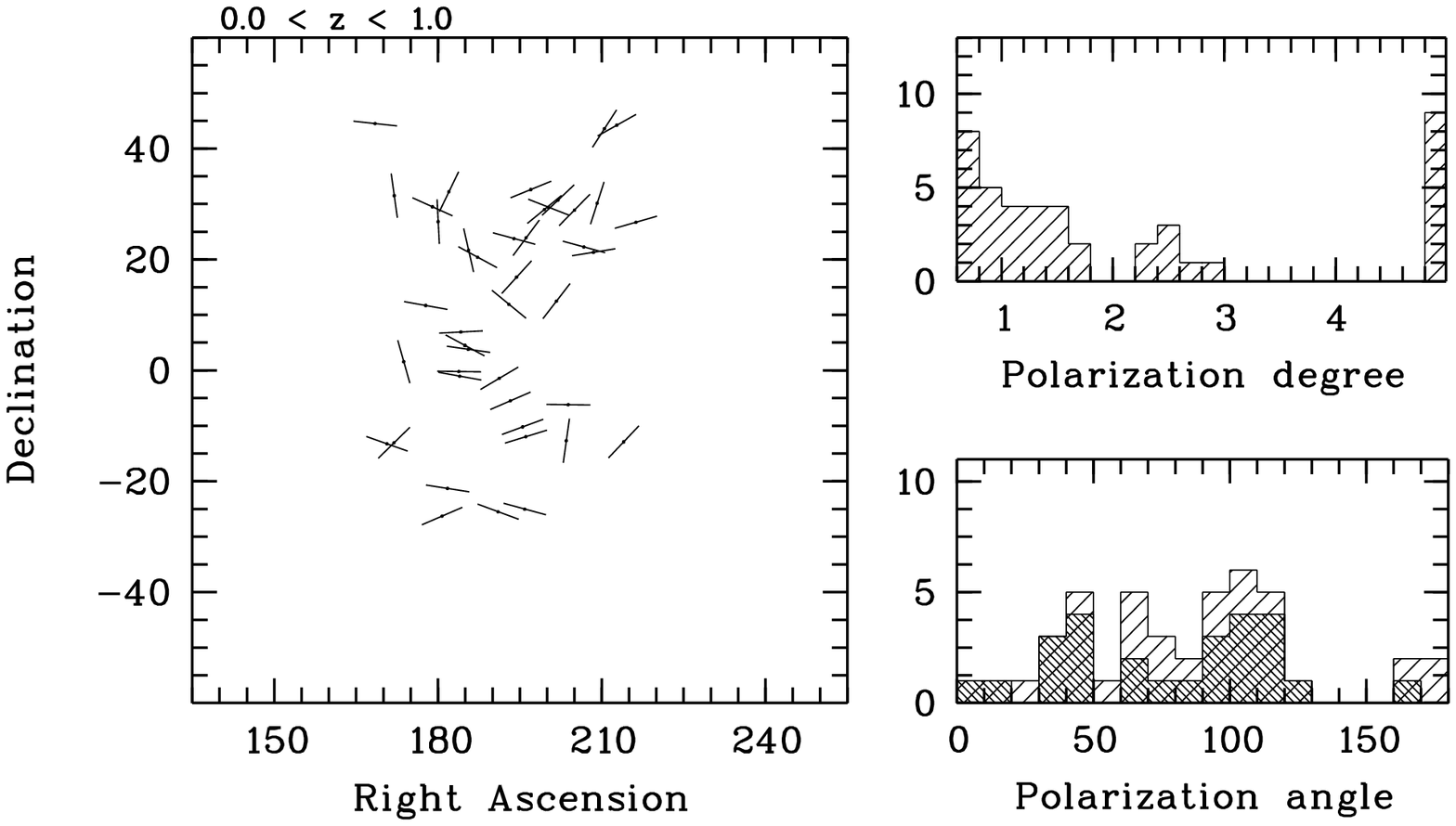}}\\[0.4cm]
\resizebox{\hsize}{!}{\includegraphics*{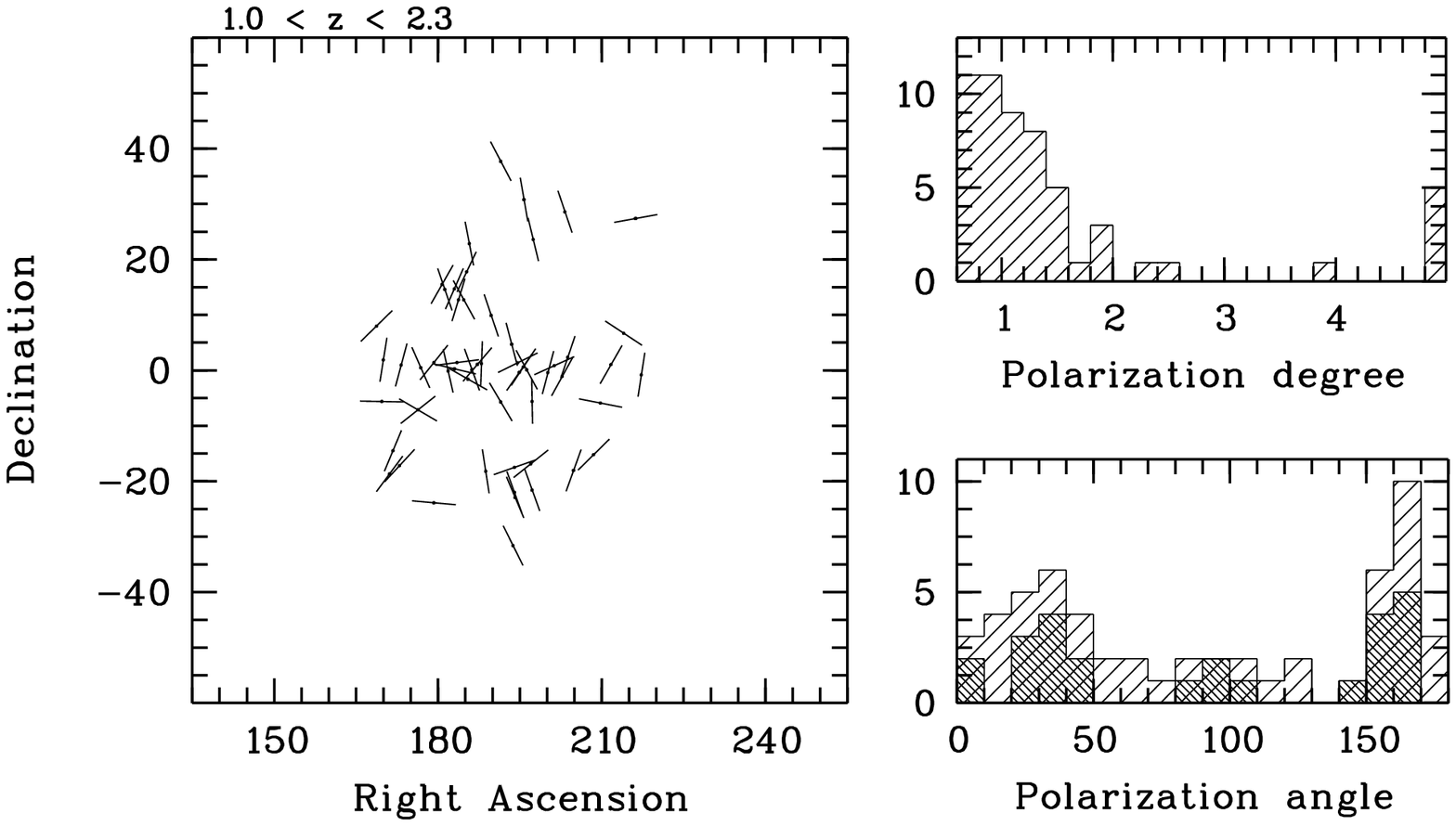}}
\caption{Maps of quasar polarization vectors in the NGP region,
together with the corresponding distributions of polarization degree
and angle. The regions illustrated are delimited in right ascension
and declination by $168\degr \leq \alpha \leq 218\degr$ and $\delta
\leq 50\degr$, and in redshift by $0.0 \leq z < 1.0$ (top, 43 objects)
and $1.0 \leq z \leq 2.3$ (bottom, 56 objects; region A1). The darker
polarization angle histograms refer to quasars with $p \geq 1.2\%$,
that is 2 times higher than the cutoff $p \geq 0.6\%$ adopted for the
full sample.}
\label{fig:mapn}
\end{figure}

\begin{figure}[t]
\resizebox{\hsize}{!}{\includegraphics*{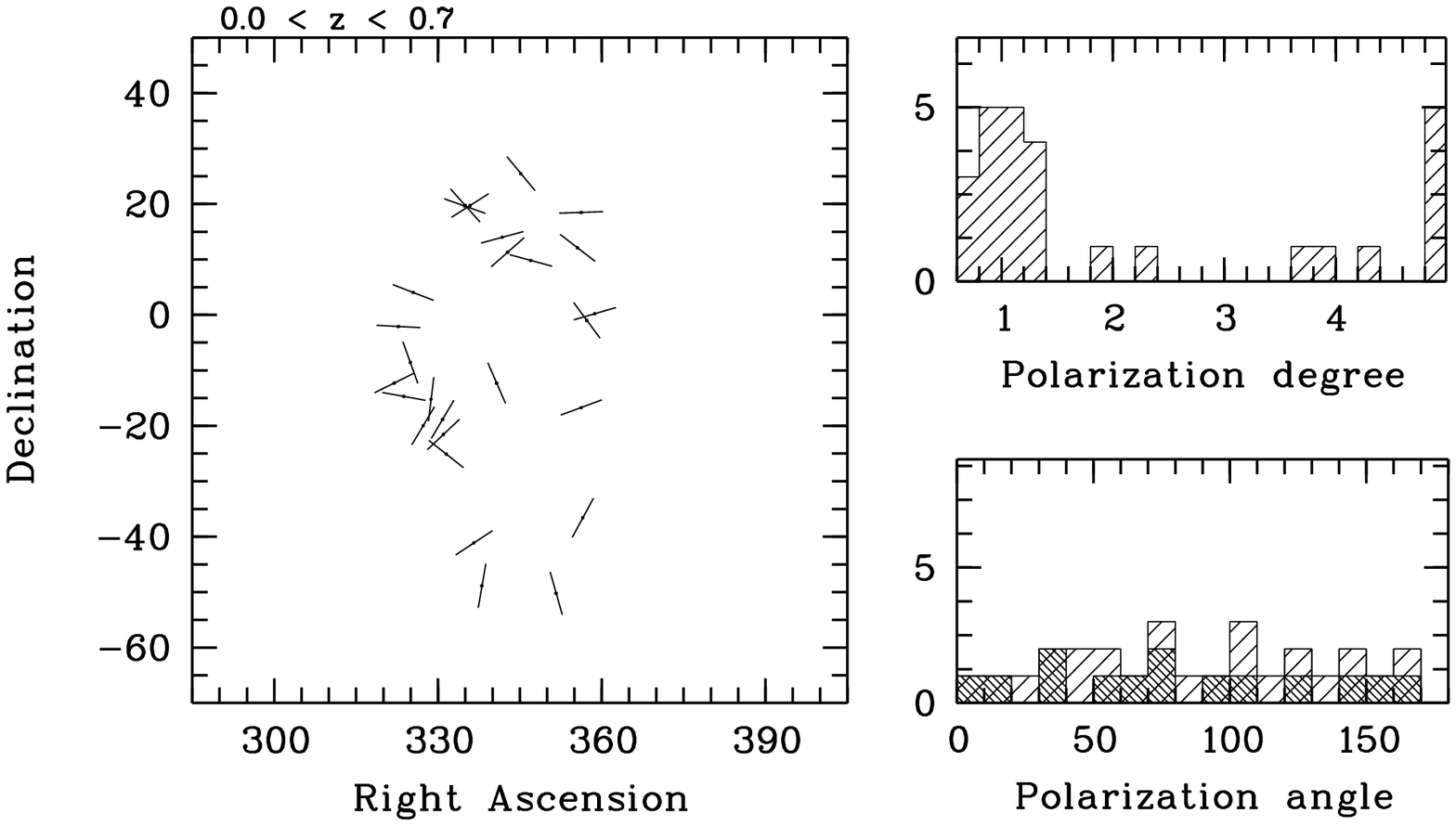}}\\[0.4cm]
\resizebox{\hsize}{!}{\includegraphics*{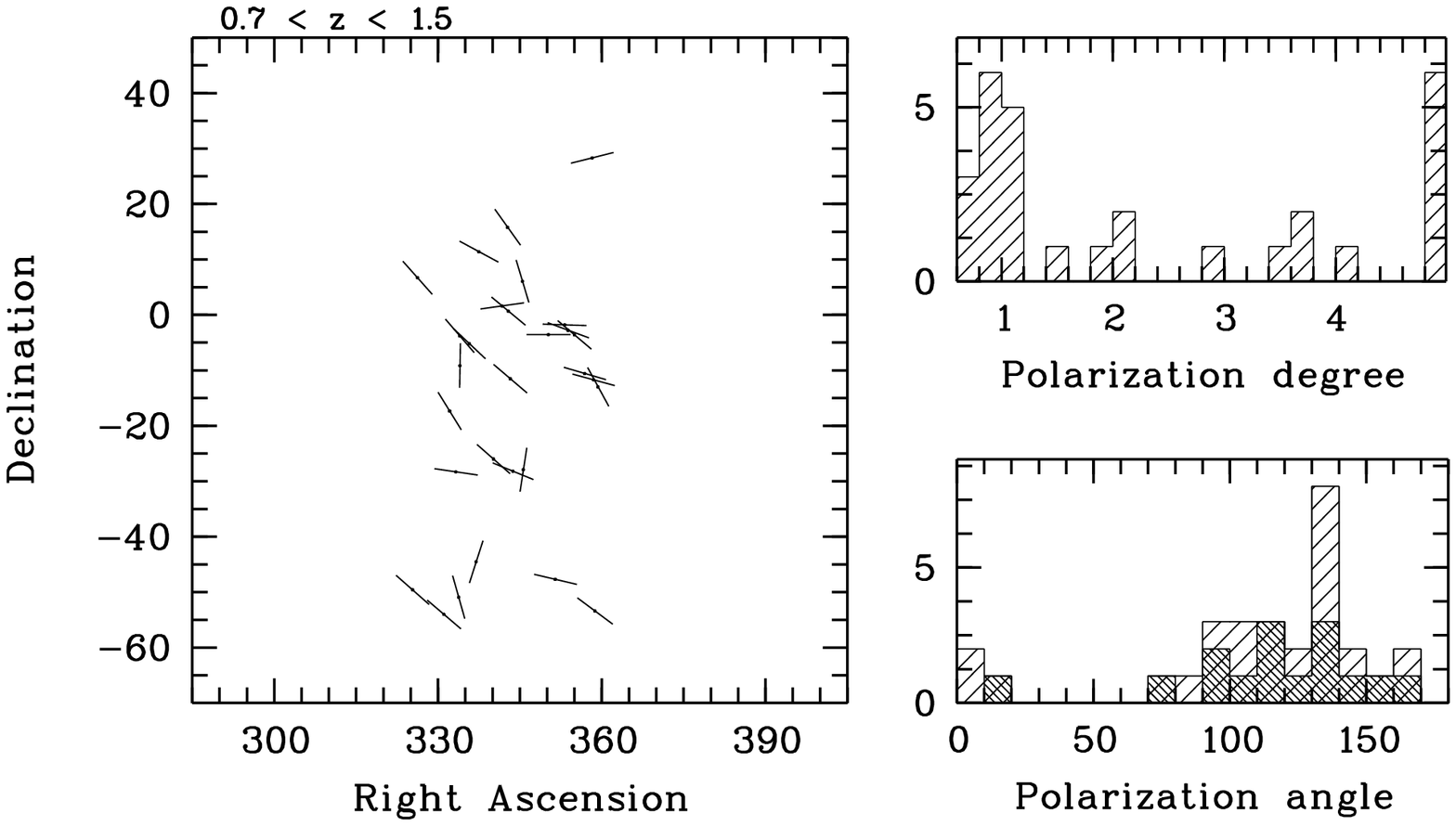}}\\[0.4cm]
\resizebox{\hsize}{!}{\includegraphics*{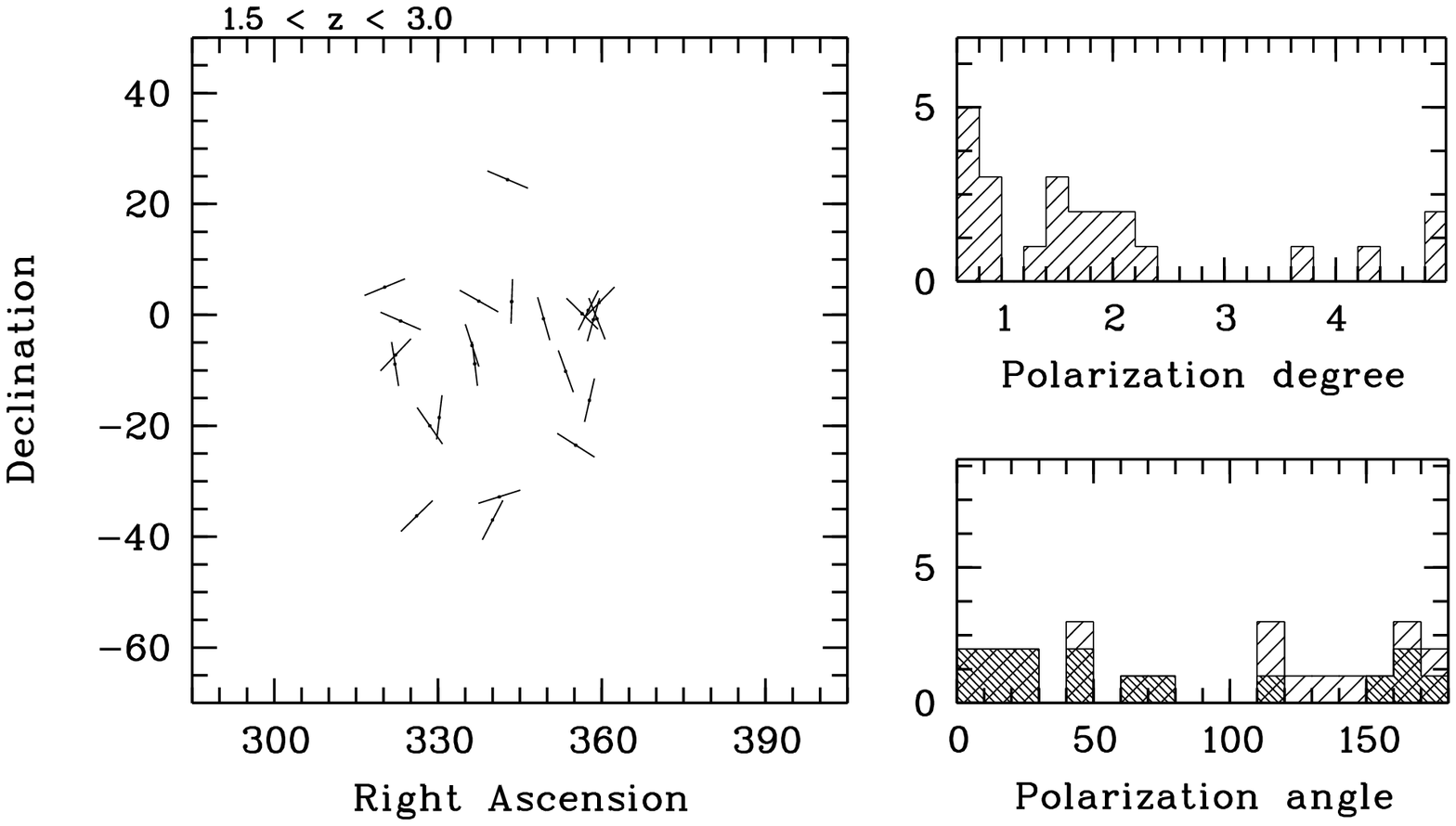}}
\caption{Maps of quasar polarization vectors in the SGP region,
together with the corresponding distributions of polarization degree
and angle. The regions illustrated are delimited in right ascension by
$320\degr \leq \alpha \leq 360\degr$, and in redshift by $0.0 \leq z
< 0.7$ (top, 27 objects), $0.7 \leq z \leq 1.5$ (middle, 29
objects; region A3) and $1.5 < z \leq 3.0$ (bottom, 23
objects). The darker polarization angle histograms refer to quasars
with $p \geq 1.2\%$. Right ascensions should be read modulo 360\degr.}
\label{fig:maps}
\end{figure}

In Fig.~\ref{fig:mapn} and~\ref{fig:maps}, we illustrate the regions
where the quasar polarization vector alignments are the most
significant. As already discussed, the borders of these regions are
not clearcut.  This is especially true in the SGP region where several
quasars with right ascensions between 0\degr\ and 40\degr\ seem to
have their polarization vectors aligned too, at least in some redshift
ranges.  However, because the spatial sampling is still poor out of
the line of sight to regions A1 and A3, we choose to essentially stick
to the limits adopted in Paper~I.

Towards the NGP (Fig.~\ref{fig:mapn}), polarization vector alignments
are seen for both the low and high redshift samples\footnote{The low
redshift region was not analysed in Sect.~\ref{ssec:statloc} because
it was not defined a priori (it overlaps but differs from the region
A2 defined in Paper~I). It is nevertheless interesting to note that
out of 43 polarized quasars in that region, 35 have their polarization
angle in the range [30\degr -- 120\degr]
(cf. Fig.~\ref{fig:mapn}). This corresponds to $P_{\rm bin} = 2.1 \,
10^{-5}$. With $p\geq1\%$ ($p\geq2\%$), 24 (13) quasars out of 30 (16)
have their polarization angle in that range and $P_{\rm bin} = 7.2 \,
10^{-4}$ ($1.1 \, 10^{-2}$).}. The average directions are definitely
different: $\overline{\theta} \simeq 79\degr$ at low-$z$ and
$\overline{\theta} \simeq 8\degr$ at high-$z$ (with $P_{\srm \!  HP} =
3\, 10^{-3}$ and $P_{\srm \! HP} = 2\, 10^{-3}$, respectively).  The
alignment in the SGP region A3 ($0.7 \leq z \leq 1.5$) is also clearly
seen, including for the higher polarization objects
(Fig.~\ref{fig:maps}).  The preferred direction is $\overline{\theta}
\simeq 128\degr$ ($P_{\srm \! HP} = 6\, 10^{-5}$).  No significant
departure to random orientations is seen in the lower or the higher
redshift SGP regions. One might suspect in the high-$z$ region an
alignment with a preferred direction different from the mid-$z$ one,
but it is not significant. However, if we only consider the 15
high-$z$ objects with $p \geq 1.2\%$, we have a weak detection with a
preferred direction $\overline{\theta} \sim 15\degr$ ($P_{\srm \! HP}
= 3\, 10^{-2}$). More data are clearly needed towards this region of
the SGP.

It is important to emphasize that, in both the NGP and the SGP
regions, the polarization degree distributions in the different
redshift sub-samples do not significantly differ (as verified with
two-sample Kolmogorov-Smirnov tests), and that both the lower and
higher polarization quasars follow the same trends.

Finally, since regions A1 and A3 are roughly opposite on the sky, we
will refer in the following to the regions defined in right ascension
and declination as in Figs.~\ref{fig:mapn} and \ref{fig:maps} as to
the ``A1--A3 axis''.

\section {Contamination by interstellar polarization in our Galaxy}
\label{sec:pism}

\begin{figure}[t]
\resizebox{\hsize}{!}{\includegraphics*{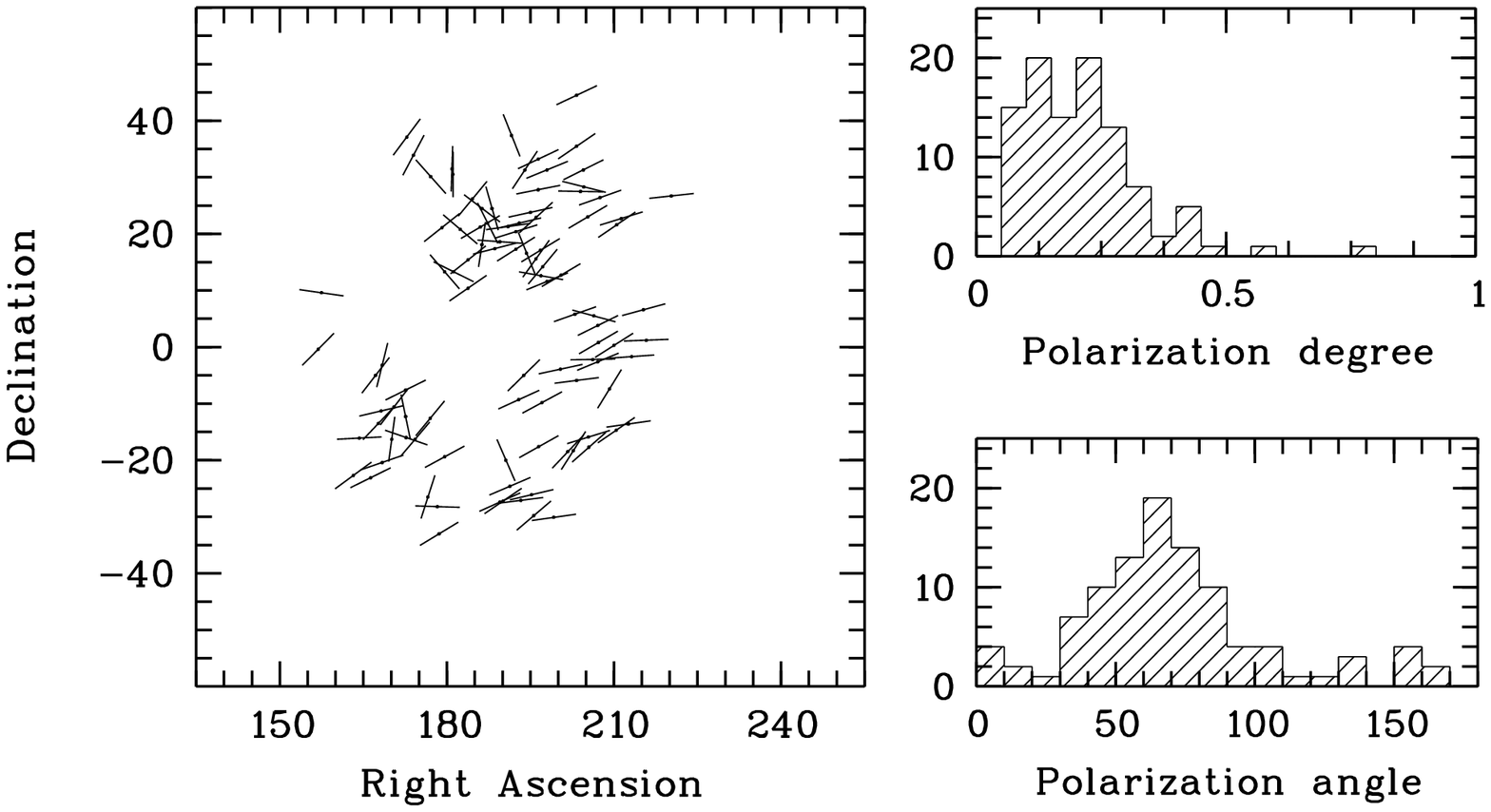}}\\[0.4cm]
\resizebox{\hsize}{!}{\includegraphics*{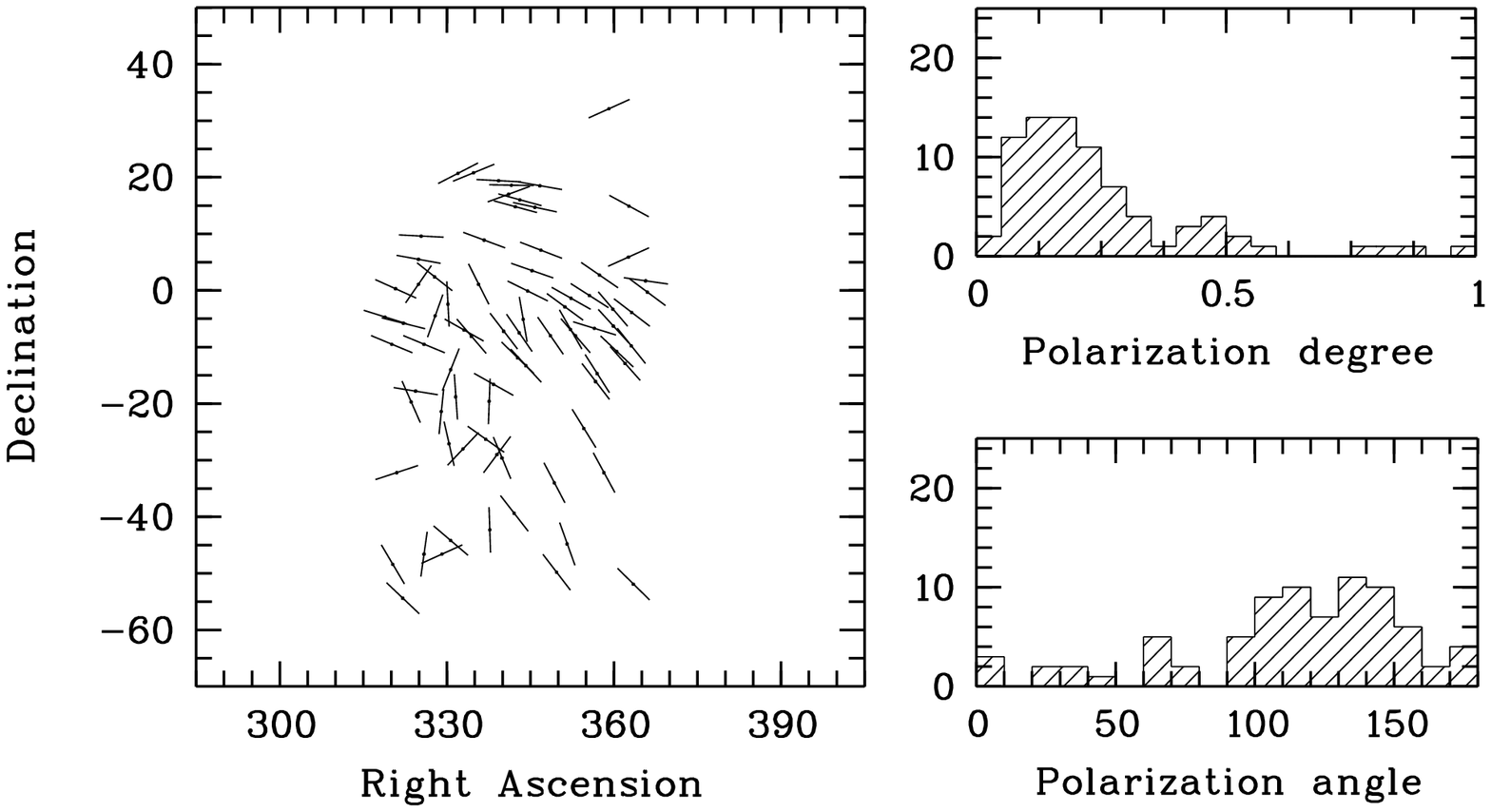}}
\caption{Maps and corresponding distributions of the interstellar
polarization measured from stars matching at best the positions of the
quasars illustrated in Fig.~\ref{fig:mapn} \& \ref{fig:maps} (top: NGP
region; bottom: SGP region). Only stars at distances $ d_{\star}\geq $
100 pc and with a polarization angle uncertainty
$\sigma_{\theta_{\star}} \leq 14\degr$ are considered. Since
$\sigma_{\theta_{\star}} = 28\fdg65 \; \sigma_{p_{\star}} /
p_{\star}$, the latter condition discards stars with $p_{\star}\simeq
0$.  The NGP is located at $\alpha=192\degr$, $\delta=27\degr$ and the
SGP at $\alpha=12\degr$, $\delta=-27\degr$}
\label{fig:mapism}
\end{figure}

The linear dichroism of aligned interstellar dust grains in our Galaxy
produces linear polarization along the line of sight which
contaminates to some extent the quasar measurements and may change
their polarization angles. Specifically, are the observed alignments
due to polarization in our Galaxy?  Although this important issue was
extensively discussed in Paper~I, it is worth to come back on it given
our larger sample.

Let us first recall that, whenever possible, we have measured the
polarization of field stars located very close to the quasars, on the
same CCD frames.  If we assume that the field star polarization
correctly represents the interstellar polarization affecting more
distant objects, then interstellar polarization in our Galaxy was
shown to have little effect on the polarization angle distribution of
significantly polarized ($p \geq 0.6\%$) quasars (Sluse et
al. \cite{SLU05}).

Since accurate field star measurements are not available for every
quasar in the sample, we consider in the following the polarization
data collected by Heiles~(\cite{HEI00}) for more than 9000 stars.  Our
field star polarization measurements are in excellent agreement with
these data (Sluse et al.~\cite{SLU05}). Fig.~\ref{fig:mapism}
illustrates polarization maps and distributions for the stars best
matching the positions of the quasars represented in
Fig.~\ref{fig:mapn} and~\ref{fig:maps}. For each quasar, we plot the
angularly closest star on the sky located at a heliocentric distance
$d_{\star} \geq$ 100~pc and with an uncertainty on the polarization
angle $\sigma_{\theta_{\star}} \leq 14\degr$; if this star is already
used, we plot the second nearest, etc, making sure that all stars are
different. Ideally one should use the most distant stars.  However, if
we increase the minimum stellar distance, the number density of stars
in the catalogue strongly decreases and the mean angular distance to
the quasars becomes larger. To keep stars within a few degrees from
the quasars, we adopt $d_{\star} \geq$ 100~pc as a good compromise. In
fact, choosing higher distance cutoffs has little effect on the
polarization angle distributions; only the polarization degrees are
slightly shifted towards higher values when more distant stars are
used.  As seen in Fig.~\ref{fig:mapism}, the polarization angles are
clearly concentrated around two preferred directions:
$\overline{\theta}_{\star} \simeq 64\degr$ in the NGP region and
$\overline{\theta}_{\star} \simeq 128\degr$ in the SGP region.  These
mean directions are typical of high galactic latitude regions of the
sky (Berdyugin et al. \cite{BER04}, Sluse et al. \cite{SLU05}) and do
not critically depend on the $\alpha$, $\delta$ bounds.

\begin{figure}[t]
\resizebox{\hsize}{!}{\includegraphics*{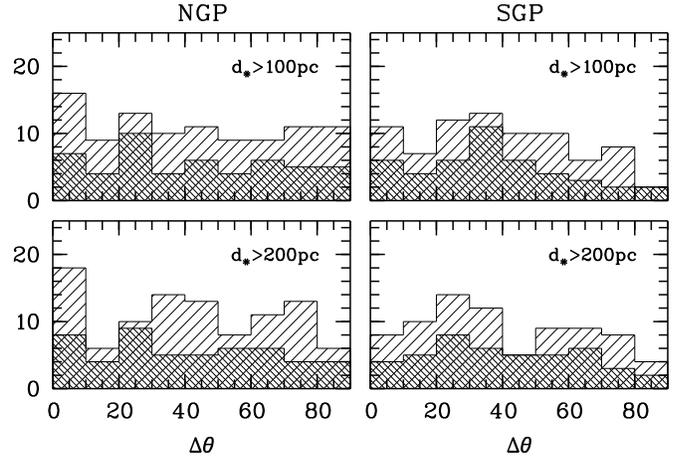}}
\caption{Distributions of the acute angle $\Delta\theta$ between the
polarization position angles of quasars and nearest stars.  The
objects represented are those considered in
Figs.~\ref{fig:mapn}--\ref{fig:mapism}. The regions of the sky towards
the NGP and the SGP are shown separately. Two cutoff in stellar
distances are considered: $d_{\star} \geq$ 100 pc and $d_{\star} \geq$
200 pc. Darker histograms illustrate distributions of $\Delta\theta$
including only quasars with $p \geq 1.2\%$.}
\label{fig:dteta}
\end{figure}

To compare quasar and stellar polarization angles, we have first
computed the difference $\Delta\theta$ between the polarization
position angles of a quasar and its nearest star: $\Delta\theta =
90\degr-|90\degr - |\theta - \theta_{\star}||$, where $\theta$ refers
to the quasar polarization angle and $\theta_{\star}$ to the stellar
one.  Distributions of $\Delta\theta$ are illustrated in
Fig.~\ref{fig:dteta}.  If quasar polarization vectors are aligned
according to interstellar polarization, one may expect a strong
clustering at small $\Delta\theta$. Such a clustering is not observed,
indicating the absence of significant correlations between quasar and
interstellar polarizations (only a weak 2$\, \sigma$ deviation is seen
in the first bin of one of the histograms).

We can also directly compare the trends seen in Fig.~\ref{fig:mapism}
to the quasar polarization vector alignments observed in
Figs.~\ref{fig:mapn} and~\ref{fig:maps}.  Towards the NGP, the
orientation of the alignment in the high-redshift region A1 appears
completely different from the direction of the interstellar
polarization.  But, on the contrary, the mean direction of the lower
redshift alignment is rather similar to that one of the interstellar
polarization, suggesting that it might be due to polarization by dust
grains in our Galaxy, although the distributions somewhat differ and
more particularly the peak seen at $\theta \simeq 110\degr$ in the
quasar polarization angle distribution.  Simple simulations show that,
apart from this peak, the clustering in the distribution of low-$z$
quasar polarization angles can be corrected by subtracting a strong
(mean $p_\star \simeq 0.7\%$) interstellar polarization at
$\overline{\theta}_{\star} \simeq 64\degr$.  The fact that higher than
observed interstellar polarization is needed to randomize the quasar
polarization angles is not supported by the observations of distant
stars (Berdyugin et al. \cite{BER04}) nor by the polarization
measurements of (a few) field galaxies (Sluse et al. \cite{SLU05}).
However, it cannot be rejected since little is known on the
interstellar polarization of very distant objects.  The fact that
interstellar polarization could be at the origin of the low-$z$
alignment is nevertheless difficult to understand since low and high
redshift quasars are located on similar lines of sight and then must
suffer the same interstellar polarization, at least on average.  One
might argue that low redshift quasars are systematically less
polarized than high redshift ones and then more affected by
interstellar polarization. But this interpretation is ruled out by the
fact that polarization degrees do not differ in the low and high
redshift quasar samples\footnote{There are several reasons which could
have explained such a difference and worth to keep in mind.  The first
one is that different types of quasars dominate the low and high
redshift sub-samples. Indeed, BAL QSOs are rarely detected at $z \leq
1.3$ such that there are proportionally more BAL quasars at high-$z$
than at low-$z$. The reverse is true for radio-loud quasars and the
fact that strongly polarized ($p \geq 5\%$) quasars are more often
found among radio-loud objects is marginally seen in the low-$z$
distribution of the polarization degree (Figs.~\ref{fig:mapn} \&
\ref{fig:maps}).  Another reason is the fact that, when measuring the
polarization either through a given filter or in white light, one
samples a bluer region of the quasar rest-frame spectrum for high
redshift objects than for low redshift ones. A wavelength dependent
quasar polarization would then also appear redshift dependent.}
(Fig.~\ref{fig:mapn}).  Furthermore, highly polarized quasars 
follow the low-$z$ alignment and low polarization ones follow the
high-$z$ alignment.  It should be emphasized that very highly
polarized quasars do follow the low-$z$ alignment: for example, out of
the 7 low-$z$ quasars with $p\geq 7 \%$, 6 have $30\degr \leq \theta
\leq 120 \degr$.  A similar behavior is observed towards the SGP
(Fig.~\ref{fig:maps}).  The mean orientation of the polarization
alignment seen for intermediate redshift quasars coincides with the
direction of the interstellar polarization in the SGP, while objects
at lower or higher redshifts on the same line of sight show
essentially random polarization angle distributions (even weakly
oriented at a different $\overline{\theta}$ at high-$z$).  Any
correction randomizing the mid-$z$ polarization angle distribution
induces a reverse concentration in the distributions of low and
high-$z$ quasar polarization angles.  And, again, the polarization
degrees do not depend on redshift, and the quasars with higher
polarization do follow the mid-$z$ alignment (Table~\ref{tab:statbin}).

Finally, we have considered pairs of quasars, i.e. quasars at small
angular distances from each other, independently of their redshift.
If interstellar polarization dominates, both quasars should be
similarly affected such that the acute angle $\Delta\theta$ between
their polarization angles is expected to cluster at small
$\Delta\theta$.  Using angular distances less than 1\degr\ and 2\degr\
(smaller distances result in too few quasar pairs), we find no
concentration at small $\Delta\theta$, either in the full sample or in
the A1--A3 region, or when only considering the low-polarization ($p
\leq 2\% $) objects more likely to be affected by interstellar
polarization (to fix the ideas, with $p \leq 2\% $, there are 17
quasar pairs with angular separations $\leq$ 1\degr\ and 61 pairs with
angular separations $\leq$ 2~\degr\ among the full sample of 355
quasars).

It is therefore very implausible that interstellar polarization is at
the origin of the observed polarization vector alignments. Most
probably, there is a small --normal-- contamination by interstellar
polarization, compatible with field star measurements, and which
possibly slightly enhances those intrinsic alignments having a similar
orientation. Given that the alignments are characterized by a broad
range of polarization angles around a preferred direction, the chance
for a coincidence is not small, especially if the mean quasar
polarization angle rotates as a function redshift as shown in the next
section.  The fact that, when cutting at $p \geq 1.2 \%$ in
Figs.~\ref{fig:mapn} and~\ref{fig:maps}, quasars with $\theta \simeq
64\degr$ ($\theta \simeq 128\degr$) are preferentially removed in the
low-$z$ NGP region (SGP region) possibly supports this view (cf. also
the first bin at $\Delta\theta \leq 10\degr$ in Fig.~\ref{fig:dteta}).
Spectropolarimetric data are in agreement with this interpretation.
Indeed, while a significant contamination by interstellar
polarization would produce a definite rotation of the polarization
angle as a function of wavelength, quasars usually show polarization
angles constant (i.e. within a few degrees) with wavelength, including
objects located in the regions of alignments (Impey et
al. \cite{IMP95}, Ogle et al. \cite{OGL99}, Schmidt \& Smith
\cite{SCH00}, Smith et al. \cite{SMI03}, Kishimoto et
al. \cite{KIS04}).  In the few quasars for which such a rotation is
observed, corrections to polarization angles do not exceed 10\degr\
(Kishimoto et al. \cite{KIS04}).

In conclusion, interstellar polarization can definitely not explain
the polarization vector alignments seen towards the NGP and more
particularly that one observed in the high redshift region A1. Towards
the SGP, it is also unlikely that interstellar polarization is at the
origin of the observed alignment, but, given the unusual nature of the
effect, more data are needed for a definite proof, namely by observing
quasars at redshifts $\geq 1.5$ where a different orientation is
suspected.

For the sake of completeness, it should be noticed that interstellar
dust grains are also linearly birefringent, such that the interstellar
medium can be seen as a weak wave-plate (Martin \cite{MAR74}, Lucas
\cite{LUC03}).  Should quasars be circularly polarized, the
interstellar medium may, under some circumstances, align their
polarization vectors along a mean direction offsetted with respect to
that one of a purely dichroic interstellar medium due to the
conversion of circular polarization into linear polarization. While
quite appealing, this mechanism cannot explain the quasar polarization
vector alignments. Indeed, the retardance is very small, roughly two
orders of magnitude smaller than that of a quarter-wave plate (Martin
\cite{MAR72}). Also, quasars are not or very weakly circularly
polarized (Landstreet \& Angel \cite{LAN72}, Impey et
al. \cite{IMP95}), including a few objects belonging to the regions of
alignment A1 and A3. And, finally, should this effect produce the
alignments, it would imply either left-handed or right-handed circular
polarization for most quasars in a given region of alignment,
i.e. still a high degree of organization on very large spatial scales.

\section {Characterizing the alignment effect}
\label{sec:prop}

In this section, we explore some characteristics of the alignment
effect with the goal to empirically derive constraints on possible
interpretations. We first focus on the redshift dependence of the
alignments. Then, we investigate whether quasars with aligned
polarization vectors are located along a preferential axis, or
not. Finally, we discuss correlations with quasar intrinsic
properties.

\subsection{The redshift dependence of the alignment effect}
\label{ssec:tetaz}

\subsubsection{Regularly spaced alignments?}
\label{sssec:regu}

\begin{figure}[t]
\resizebox{\hsize}{!}{\includegraphics*{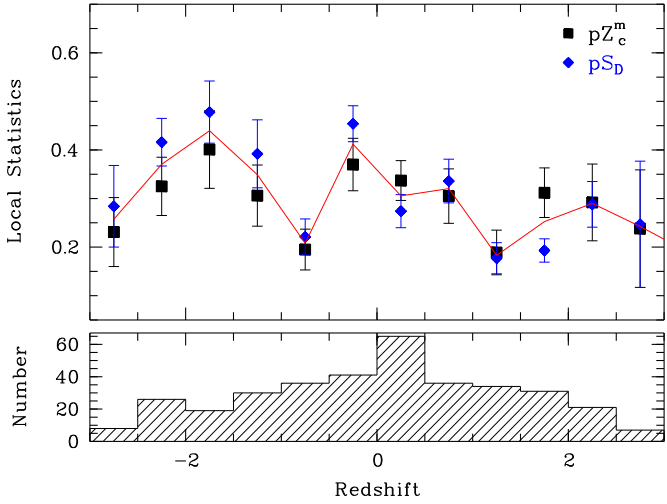}}
\caption{The local statistics $s_i$ of the $pS_{\sit D}$ and $pZ_c^m$
tests computed with $\nv$ = 40 are represented here as a function of
the redshift. The full sample of 355 quasars is considered. The $s_i$,
which have been multiplied by a constant factor for convenience, are
averaged over redshift bins $\Delta z = 0.5$; the error bars represent
the uncertainty of the mean.  The highest values of $s_i$ indicate the
strongest departures to uniform distributions of polarization angles.
The line joins the mean $s_i$ values from the $pS_{\sit D}$ and $pZ_c^m$
tests. Since the $s_i$ are computed for $\nv$ = 40, the data points
are not independent.  Redshifts are counted positively for objects
located in the North Galactic Cap and negatively for those ones in the
South Galactic Cap.  The histograms give the number of quasars in each
redshift bin. }
\label{fig:statz1}
\end{figure}

When computing the global statistics in Sect.~\ref{ssec:statglo}, a
local statistic $S_i$ is defined for each object $i$ and its $\nv$
neighbours.  It is evaluated for the original data, $S_i^{\star}$, as
well as for every simulation. We may then calculate $<\!  S_i \! >$,
the average over the whole set of simulations, and $\sigma_i$, the
corresponding standard deviation, such that the quantity $s_i \propto
|\! < \! S_i \! > - S_i^{\star}| \, / \sigma_i$ provides a measure of the
local departure to an uniform distribution of polarization
angles. For the $S$-type tests, only small values of $S_i^{\star}$
indicate coherent orientations and $s_i$ is set to zero when
$S_i^{\star}$ is larger than $<\!  S_i \! >$. For the $Z$-type tests,
$s_i$ is set to zero when $S_i^{\star}$ is smaller than $<\!  S_i \!
>$ (cf. Paper~I for details).

In Fig.~\ref{fig:statz1}, we plot the quantity $s_i$, averaged over
redshift bins, as a function of the redshift. The full sample of 355
quasars is considered.  $s_i$ is computed from the $pS_{\sit D}$ and
$pZ_c^m$ tests with $\nv$ = 40. The $S_{\sit D}$ and $Z_c^m$ tests,
not represented here, give similar results. For both statistical
tests, the run of $s_i$ with redshift shows a cyclic behavior
suggesting a regular alternance of regions of aligned and randomly
oriented polarization vectors. The minima at $z \simeq 0.7$ towards
the SGP and $z \simeq 1.2$ towards the NGP correspond to transition
redshifts discussed in previous sections.  It must be emphasized that
adjacent data points are not independent due to the fact that the
statistics $s_i$ are evaluated using $\nv = 40$ nearest neighbours.

\begin{figure}[t]
\resizebox{\hsize}{!}{\includegraphics*{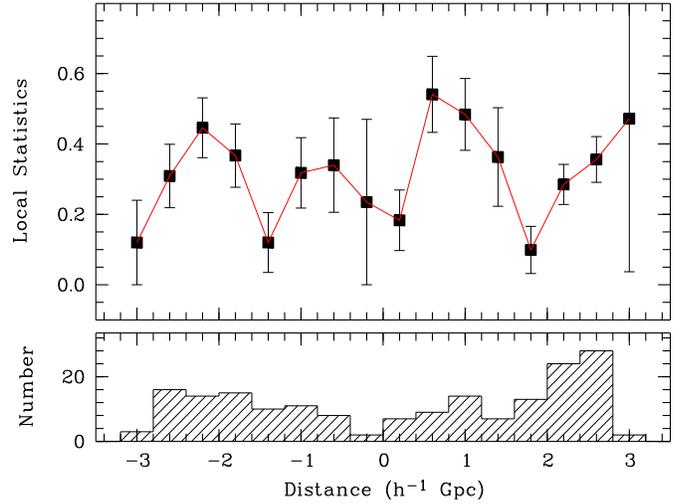}}
\caption{Same as Fig.~\ref{fig:statz1}, except that a comoving
distance scale is used and that only those quasars along the A1--A3
axis are considered.  Redshifts are transformed into comoving
distances using $r\, = 6 \, (1 - (1+z)^{-1/2}) \,\; h^{-1}$ Gpc, where
$h$ is the Hubble constant in units of 100 km s$^{-1}$ Mpc$^{-1}$.
Bin size is $\Delta r = 0.4 \, h^{-1}$ Gpc. Distances are counted
positively for objects located in the North Galactic Cap and
negatively for those ones in the South Galactic Cap.}
\label{fig:statz2}
\end{figure}

The redshift dependence of the alignment effect is best seen in
Fig.~\ref{fig:statz2}, when only the $pZ_c^m$ test and those quasars
along the A1--A3 axis (as defined in Sect.~\ref{sec:maps}) are
considered.  A comoving distance scale is used to emphasize the
regular variation of the alignment effect with cosmological
distance. This variation appears quasi-periodic, the distance between
two extrema being $\sim$ 1.5 $h^{-1}$ Gpc. Such a behavior may clearly
constitute an important clue to the interpretation of the alignment
effect (Sect.~\ref{sec:inter}).  Additional data at high redshift are
needed to confirm it.  Interestingly enough, a quasi-periodicity in
quasar polarization vector alignments, if correctly understood, may
potentially constitute a new distance indicator.

\subsubsection{Rotation of the mean direction with redshift?}
\label{sssec:rota}

\begin{figure}[t]
\resizebox{\hsize}{!}{\includegraphics*{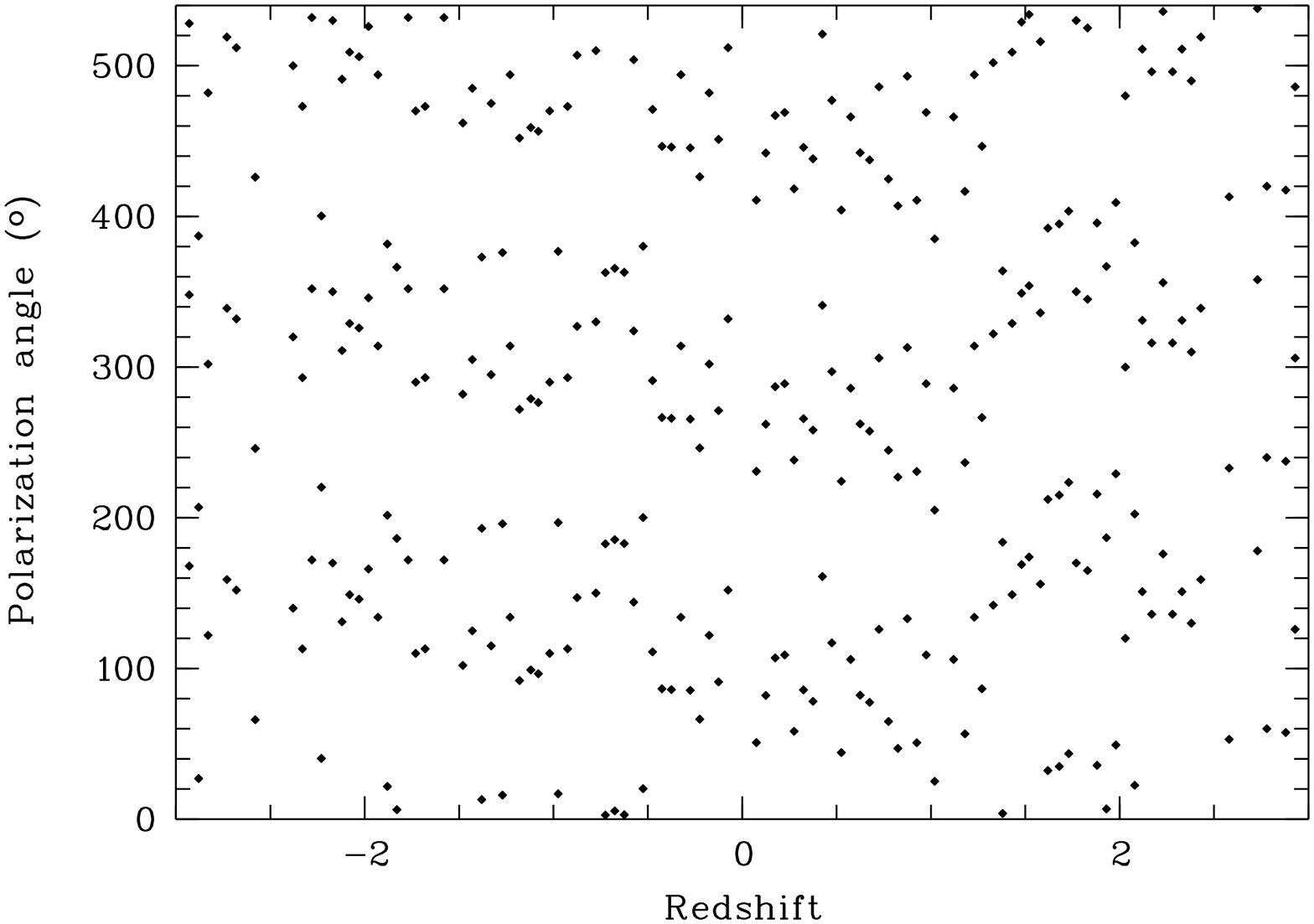}}
\caption{The quasar polarization angles as a function of the
redshift.  Redshifts are counted positively for objects located in
the North Galactic Cap and negatively for those ones in the South
Galactic Cap. Polarization angles are vectorially averaged over
redshift bins $\Delta z = 0.05$.  The full sample of 355 quasars is
used.  To facilitate detection of patterns, each data point ($z$,
$\theta$) is replicated at ($z$, $\theta$+180\degr) and ($z$,
$\theta$+360\degr).}
\label{fig:tetaz1}
\end{figure}

Results presented in Sect.~\ref{sec:stats} \& \ref{sec:maps} also
indicate that the mean polarization angle of quasars changes with
redshift. In Fig.~\ref{fig:tetaz1} we plot the polarization angles of
the 355 quasars, slightly averaged over redshift bins, as a function
of the redshift. To emphasize possible relationships, each data point
is plotted three times in the graph, adding $n \times 180\degr$ to the
polarization angles, with $n = 0, 1, 2$. It appears quite clearly that
the polarization angles are not randomly distributed over
redshifts. Some patterns may be seen as, for example, a continuous
decrease of the polarization angle with increasing redshift.  The
possible relation is more ambiguous around the redshifts $z \simeq
0.7$ towards the SGP and $z\simeq 1.2$ towards the NGP, which
correspond to the redshift ranges where no alignment is detected
(Fig.~\ref{fig:statz1}). Another possible relation could be a decrease
of the polarization angle with $z$ in the SGP region ($z < 0$)
followed by an increase in the NGP region ($z > 0$).

To investigate more quantitatively possible correlations, we make use
of statistical methods which take into account the circular nature of
the data; they are described in Fisher (\cite{FIS93}). First, we map
the redshift onto the circle using $\phi = 2 \, \tan^{-1} z$, where
$z$ is taken to be negative for objects located in the South Galactic
Cap and positive for those ones in the North Galactic Cap.  As usual,
we take into account the axial nature of the polarization angles
$\theta$ by multiplying them by a factor 2. Then we analyse possible
correlations between $\phi$ and $\theta$ using the angular--angular
correlation coefficients $\widehat{\Pi}_{n}$ and
$\widehat{\rho}_{_{T}}$. $\widehat{\Pi}_{n}$ is a correlation
coefficient based on the circular ranks of the $\phi_i$ and
$\theta_i$. It assesses monotone association between $\phi$ and
$\theta$.  $\widehat{\rho}_{_{T}}$ estimates the linear association
between $\phi$ and $\theta$ based on the simple models $\phi = \theta
$ + cst or $\phi = -\theta $ + cst.  Because it is independent of the
scaling of $z$ (including its tranformation into a more physical
distance scale), the $\widehat{\Pi}_{n}$ correlation coefficient is
more general.  The hypothesis that $\phi$ and $\theta$ are independent
is rejected if $\widehat{\Pi}_{n}$ or $\widehat{\rho}_{_{T}}$ differ
too much from zero.  The probability that a value more different from
zero than the observed values of $\widehat{\Pi}_{n}$ and
$\widehat{\rho}_{_{T}}$ would occur by chance among uncorrelated
$\phi$ and $\theta$ is evaluated on the basis of 10$^{5}$
permutations, shuffling the polarization angles over the redshifts.

\begin{table}[t]
\caption{Results of correlation tests}
\label{tab:statteta}
\begin{tabular}{lrrrrrr}\hline\hline \\[-0.10in]
Sample & & & $\widehat{\rho}_{_{T}}$ \ & $P_{\rho}$ \ &  
$\widehat{\Pi}_{n}$ \ & $P_{\pi}$ \ \\
\hline \\[-0.10in]
355 \ S1 & eq & & $-$0.0254 & 3 10$^{-4}$ &  $-$0.0294  & 6 10$^{-5}$ \\
183 \ S1 & eq & & $-$0.0702 & $<$1 10$^{-5}$ &  $-$0.0770  & $<$1 10$^{-5}$ \\
129 \ S1 & eq & & $-$0.1113 & $<$1 10$^{-5}$ &  $-$0.1224  & $<$1 10$^{-5}$ \\[0.1cm]

355 \ S2 & eq & & $-$0.0288 & 5 10$^{-5}$ &  $-$0.0304  & 3 10$^{-5}$ \\
183 \ S2 & eq & & $-$0.0201 & 3 10$^{-2}$ &  $-$0.0304  & 4 10$^{-3}$ \\
129 \ S2 & eq & & $-$0.0124 & 2 10$^{-1}$ &  $-$0.0132  & 2 10$^{-1}$ \\[0.1cm]

355 \ S1 & sg & & $-$0.0236 & 3 10$^{-4}$ &  $-$0.0188  & 1 10$^{-3}$ \\
183 \ S1 & sg & & $-$0.0572 & 2 10$^{-5}$ &  $-$0.0488  & 6 10$^{-5}$ \\
129 \ S1 & sg & & $-$0.0782 & 5 10$^{-5}$ &  $-$0.0801  & 3 10$^{-5}$ \\[0.1cm]

355 \ S2 & sg & & $-$0.0202 & 9 10$^{-4}$ &  $-$0.0216  & 5 10$^{-4}$ \\
183 \ S2 & sg & & $-$0.0201 & 3 10$^{-2}$ &  $-$0.0240  & 1 10$^{-2}$ \\
129 \ S2 & sg & & $-$0.0075 & 4 10$^{-1}$ &  $-$0.0027  & 7 10$^{-1}$ \\
\hline\\[-0.2cm]\end{tabular}
\end{table}

The results of the statistical analysis are given in
Table~\ref{tab:statteta}. The full sample of 355 quasars is considered
as well as the sample of 183 objects along the A1--A3 axis. Out of
these 183 quasars, a sub-sample of 129 objects with $p \leq 2\%$ is
also considered.  Looking at Fig.~\ref{fig:tetaz1}, we have noticed
that polarization angles either continuously decrease with increasing
redshift, or decrease in the SGP region ($z < 0$) and increase in the
NGP one ($z > 0$).  Both possibilities are tested by using the
polarization angles ``as measured'' in both the SGP and NGP regions
(case S1), or by taking $180\degr -\theta$ instead of $\theta$ for
those objects located in the NGP region (case S2).  Since the mean
direction of the alignment in region A1 was found to be roughly
parallel to the supergalactic plane (Paper~II), the Local Supercluster
may constitute a more natural reference frame.  We then run the tests
with the polarization angles expressed in both the equatorial (eq) and
supergalactic (sg) coordinate systems.  It essentially appears from
Table~\ref{tab:statteta} that the correlation between quasar
polarization angles and redshifts is very significant, especially in
the case S1 and when only those quasars of the A1--A3 axis are
considered. The correlation is significant in both coordinate systems.
The tests were also carried out for the sample of 172 quasars obtained
when removing the objects which belong to the A1--A3 regions. No
correlation was found in that case, again suggesting that the observed
effect is mainly due to the objects along the A1--A3 axis.

\begin{figure}[t]
\resizebox{\hsize}{!}{\includegraphics*{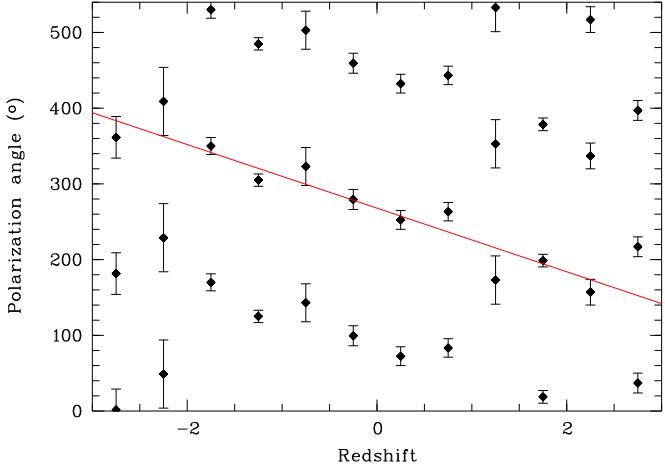}}
\caption{The quasar polarization angles, vectorially averaged over
redshift bins $\Delta z=0.5$, as a function of the redshift.  Redshifts
are counted positively for objects located in the North Galactic Cap
and negatively for those ones in the South Galactic Cap.  Only the 183
quasars belonging to the A1--A3 axis are considered here.  Error bars
represent 68\% angular confidence intervals for the circular mean
(Fisher \cite{FIS93}); they must be seen with caution when the number
of quasars per redshift bin is small, i.e. at large $z$. Errors are
higher at $z \simeq -0.7$ and $z \simeq 1.2$ in agreement with
previous results.  Data points are replicated at ($z$, $\theta$), at
($z$, $\theta$+180\degr) and at ($z$, $\theta$+360\degr) as in
Fig.~\ref{fig:tetaz1}. The fitted line is given by $\bar{\theta} =
268\degr - 42\degr \, z$ (see text).}
\label{fig:tetaz2}
\end{figure}

\begin{figure}[t]
\resizebox{\hsize}{!}{\includegraphics*{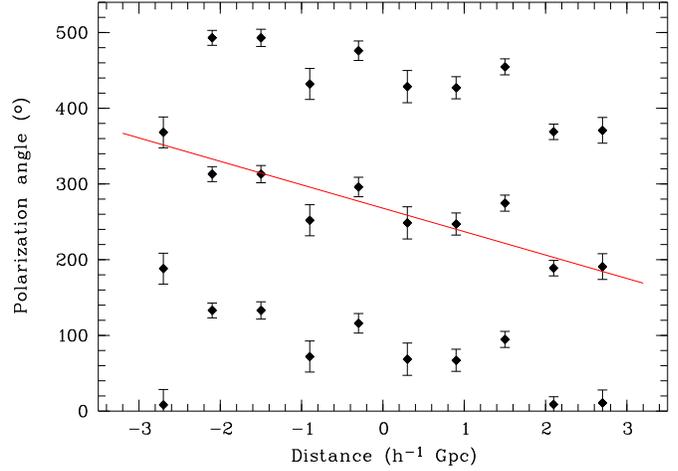}}
\caption{Same as Fig.~\ref{fig:tetaz2}, except that a comoving
distance scale is used (cf.  Fig.~\ref{fig:statz2}).  Bin size is
$\Delta r= 0.6 \, h^{-1}$ Gpc. The superimposed line is $\bar{\theta}
= 268\degr - 31\degr \, r$, where $r$ is the comoving distance in
$h^{-1}$ Gpc.}
\label{fig:tetar2}
\end{figure}

The S1 correlation is illustrated in Fig.~\ref{fig:tetaz2} for the 183
quasars along the A1--A3 axis. It shows a surprisingly clear
quasi-linear relation (which is even better defined for the sub-sample
of 129 quasars with $p \leq 2\%$, in agreement with the results of
Table~\ref{tab:statteta}).  A simple linear regression over the 7 most
accurate data points\footnote{Regression may also be performed using
the unbinned data set. In this case the dispersion of the polarization
angles is too large and a circular analysis is mandatory. According to
Fisher~(\cite{FIS93}), we may fit the following model to the data:
$\bar{\theta} = \bar{\theta}_{0} + \tan^{-1} \beta \, z$.  The
$\theta_i$ are assumed to be drawn from a von Mises distribution, with
no dependence of the dispersion upon redshift. The maximum likehood
estimates of $\beta$ and $\bar{\theta}_{0}$ are $-1.08 \pm 0.27$ and
$84\degr \pm 6\degr$, respectively, for the [183-S1-eq] sample. This
confirms that the correlation between quasar mean polarization angles
and redshifts is significant.  The analysis with the whole sample of
355 quasars gives similar results.}  gives $\bar{\theta} =
(88\degr\pm6\degr) - (42\degr\pm4\degr) \, z$. This relation
reproduces fairly well the preferred directions seen in
Figs.~\ref{fig:mapn} \& \ref{fig:maps}.  It corresponds to a rotation
of roughly $\pm$ 90\degr\ over the sampled redshift range. It is
important to realize that, in general, one may expect a step-like
discontinuity at $z = 0$, and then need a more complicated fitting of
the S1 correlation. This is due to the way position angles are defined
on the celestial sphere.  Let us imagine a large-scale structure
crossing the observer at $z =0$ and for which we measure a position
angle $\theta$ looking towards the NGP.  For the same structure, we
measure a position angle $-\theta$ looking towards the SGP, which
makes a discontinuity in the position angles at $z = 0$ (unless
$\theta \simeq$ 0\degr\ or 90\degr\ in the adopted coordinate system).
Furthermore, both the slope and the constant of the linear fit depend
on the coordinate system.  This problem can be partially overcome by
parallel transporting the polarization vectors at a given location.
More precisely, we may parallel transport the polarization vectors at
the position ($\alpha_c$, $\delta_c$) for those quasars located in the
NGP region and ($\alpha_c+180\degr$, $-\delta_c$) for those ones in
the SGP region. A reasonable choice is close to the A1--A3 axis, say
$\alpha_c = 180\degr$ and $\delta _c = 10\degr$.  In the S2 case, this
makes the $\widehat{\rho}_{_{T}}$ and $\widehat{\Pi}_{n}$ tests fully
coordinate invariant, as well as the slope of the linear regression
models.  In the S1 case, the tests still depend on the coordinates
through the discontinuity at $z =0$.  With parallel transport, there
is a tendency for the S1 correlation to be slightly more significant
than in Table~\ref{tab:statteta}, and for the S2 correlation to be
slightly less significant.  However, we find that the results of the
statistical tests and regressions are essentially unchanged, provided
that one parallel transports the polarization vectors close to the
A1--A3 axis.  Results are robust to small changes of ($\alpha_c$,
$\delta_c$). Finally, the rotation of the mean polarization angle is
also clearly seen when using comoving distances instead of redshifts
(Fig.~\ref{fig:tetar2}). A linear regression gives $\bar{\theta} =
(88\degr\pm5\degr) - (31\degr\pm3\degr) \, r$, where $r$ is the
comoving distance in $h^{-1}$ Gpc.

The existence of a significant continuous rotation of the mean
polarization angle as a function of the redshift
\footnote{In principle, the redshift dependence of the mean
polarization angle may also be a colour effect rather than a distance
effect because we are sampling quasar rest-frame spectra at different
wavelengths. However spectropolarimetry in the ultraviolet-visible
range of (a few) quasars along the A1--A3 axis does not show
significant rotation of the continuum polarization angle with
wavelength (Impey et al. \cite{IMP95}, Ogle et
al. \cite{OGL99}).  In general, very few quasars show a rotation
of the polarization angle with wavelength. For example, among the 28
polarized quasars studied by Ogle et al. (\cite{OGL99}), only 3
display a rotation of the polarization angle, typically $\Delta\theta
\sim$ 20\degr\ over the full ultraviolet-visible spectral range.}
and the symmetry of its dependence are clearly key properties of the
alignment effect.  While the regions of alignments may look at first
glance quite isolated, their properties appear connected on large
cosmological distances.  The fact that a rotation $\simeq 45\degr$
occurs roughly over the distance between two strong alignments ($\sim$
1.5 $h^{-1}$ Gpc, Fig.\ref{fig:statz2}) suggests that both phenomena
are probably due to a single mechanism.  The simple mirror-like (S1)
symmetry of the $\bar{\theta}$ -- $z$ relation is remarkable: rotation
is clockwise with increasing redshift in NGP hemisphere and
counter-clockwise in the SGP one.  This relation is the best defined
and the most significant, but a counter-clockwise rotation in both the
South and North Galactic Caps (S2 symmetry) cannot be totally
excluded, especially if we consider the statistical tests applied to
the full sample (Table~\ref{tab:statteta}).  Also, due to the
180\degr\ uncertainty, several other complicated or asymmetric
solutions to the $\bar{\theta}$ -- $z$ relation could be
imagined. Measurements of quasar polarization angles at redshifts $z
\geq 2.5$ are needed to extend and confirm the mirror-like symmetry of
the $\bar{\theta}$ -- $z$ relation.  Moreover, it would allow us to
know if the full rotation can exceed 90\degr\ or whether the mean
polarization angle oscillates between 0\degr\ and 90\degr.  Finally,
it is interesting to note that extrapolating the redshift dependence
of the mean polarization angle at $z \simeq 0$ gives $\bar{\theta}
\simeq 90\degr$. While this would be an unpleasant coincidence in the
equatorial coordinate system, this value corresponds to
$\bar{\theta}_{_{SG}} \simeq 0\degr$ in the supergalactic reference
frame, which means that the polarization vectors of hypothetical
quasars at $z \simeq 0$ should be aligned perpendicular to the
supergalactic plane. It is also worth to note that $\bar{\theta}$ at
$z=0$ is different from the mean directions of the interstellar
polarization (Fig.~\ref{fig:mapism}).

\subsection{Is there an alignment axis?}
\label{ssec:axis}

The fact that the most significant regions of polarization vector
alignments are roughly opposite on the sky suggests that they may
define an axis in the Universe.  The possible coincidence of such an
axis with other preferred directions in the sky may provide
important clues to the origin of the alignment effect.  For example,
region A1 is in the direction of Virgo, the center of the Local
Supercluster (see also Sect.~\ref{ssec:statglo}), and the A1--A3 axis
is not far from the direction of the Cosmic Microwave Background (CMB)
dipole ($\alpha$ = 168\degr, $\delta$ = $-$7\degr).  Recent analyses
of WMAP data indicate that several large-scale anisotropies in the CMB
are possibly related to this direction (e.g. Tegmark et
al. \cite{TEG03}).  These possible coincidences are discussed in
details by Ralston and Jain (\cite{RAS04}).

\begin{figure}[t]
\resizebox{\hsize}{!}{\includegraphics*{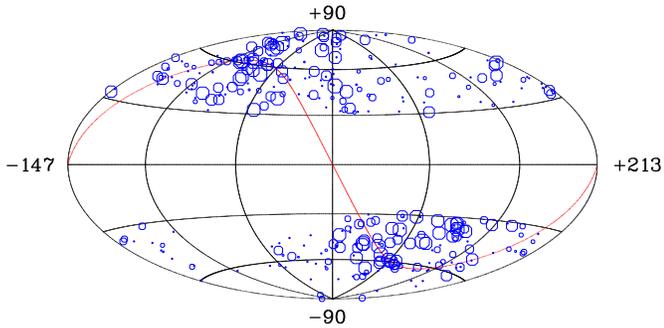}}
\caption{Hammer-Aitoff projection of the quasar positions on the sky,
in Galactic coordinates. The 355 objects are plotted. The radius of
the circles is given by $\rho_i \propto \exp s_i - 0.9$, where $s_i$
refers the statistic defined in Sect.~\ref{ssec:tetaz} for the
$pZ_c^m$ test and $\nv$ = 40; the larger the circle the more
significant the alignment at that point.  The superimposed line gives
the location of the celestial equator.}
\label{fig:axis}
\end{figure}

\begin{figure}[t]
\resizebox{\hsize}{!}{\includegraphics*{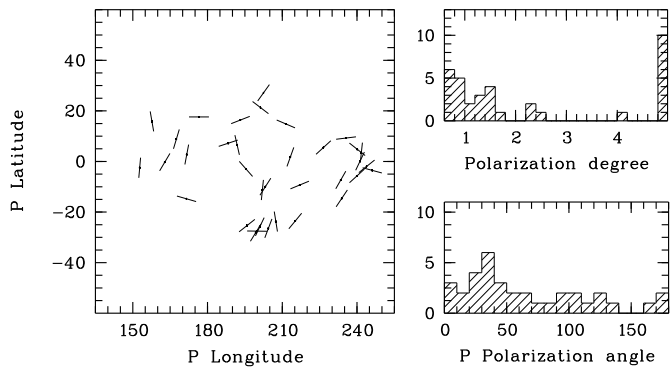}}
\caption{Maps of quasar polarization vectors and the corresponding
distributions of polarization degree and angle for the low-redshift
($z \leq 1$) objects located in the region of the sky defined in
galactic coordinates by $\bgal \geq 30\degr$ and by $35\degr \leq
\lgal \leq 175\degr$ (i.e. in the upper right quadrant of
Fig.~\ref{fig:axis}). Positions and polarization angles are projected
in a coordinate system of northern pole [$\alpha_p = 0\degr$,
$\delta_p = 30\degr$] and denoted by ``P''.  The distribution of
polarization angles is weakly coherently oriented with $P_{\srm \! HP} =
2\, 10^{-2}$ and a preferred direction of 37$\degr$ in that coordinate
system.}
\label{fig:aliout}
\end{figure}

The evidence for an alignment axis, also suggested from the maps shown
in Fig.~\ref{fig:sl_pole}, is best illustrated in Fig.~\ref{fig:axis}
where a dipole-like anisotropy is clearly seen in the distribution of
the ``most aligned'' quasars, as measured from local
statistics\footnote{Ralston \& Jain (\cite{RAS04}) computed ($\lgal
=266\degr$, $\bgal = 61\degr$) for the axis. From
Fig.~\ref{fig:sl_pole} we found ($\alpha = 180\degr$, $\delta =
10\degr$) which corresponds to ($\lgal =267\degr$, $\bgal =
69\degr$). In Galactic coordinates, the CMB dipole points towards
($\lgal =264\degr$, $\bgal = 48\degr$).}. However, this distribution
is definitely affected by observational biases. Indeed, in Paper~I, we
discovered polarization vector alignments for quasars located not far
from the celestial equator (the so-called regions A1 and A3), and we
subsequently put emphasis on these regions when gathering additional
data.  Moreover, quasars are often surveyed in equatorial fields which
provide the bulk of targets for a southern hemisphere observatory.
So, it is not unexpected that the highest quasar densities and the
highest significances do appear in these regions. The effect of such
intrincate biases on the significance of the axis is difficult to
estimate (and is clearly beyond the scope of this paper). However,
since preferred axes in the CMB are independently suggested, based on
homogeneous data samples and in agreement with the anisotropy seen in
Fig.~\ref{fig:axis}, the fact that polarization data are compatible
with a possible alignment axis is worth to keep in mind.

A related question is the following: are there really no polarization
vector alignments out of the A1--A3 axis?  Although measurements are
not very numerous, we have tried to identify possible alignments in
the upper right quadrant in Fig.~\ref{fig:axis}.  Since these objects
are close to the celestial north pole, alignments are blurred when
polarization angles are measured in equatorial coordinates.  We then
consider quasar positions and polarization angles in a coordinate
system of northern pole [$\alpha_p = 0\degr$, $\delta_p = 30\degr$].
A possible alignment is tentatively identified in
Fig.~\ref{fig:aliout}; it may constitute a high declination extension
to the low-redshift alignment seen in Fig.~\ref{fig:mapn}.  But, given
that very few objects have been measured in these regions (only 46 out
of 355 quasars belong to the third of the sky opposite to the regions
of highest significance seen in Fig.~\ref{fig:axis}), it is difficult
to conclude and the fact that significant polarization vector
alignments do or do not exist far from the A1--A3 axis is still to be
demonstrated.

\subsection{Relation to quasar intrinsic properties}

Optical polarization is known to be related to other quasar intrinsic
properties like spectral type or morphology.  In order to understand
the alignment effect, it is important to know whether these relations
are still valid for those objects in the regions of alignments.  A
full answer would require a much larger sample and more information on
the objects than available in the literature. Some questions have
nevertheless been adressed in Paper~I and~II and are summarized here.

In the high-redshift region of alignment A1, several types of quasars
have been observed, namely radio-loud, radio-quiet, and BAL
quasars. These distinctions are based on the spectral characteristics
of the objects.  First, it is important to note that polarization
vector alignments are not restricted to one category of objects.  In
region A1, BAL, non-BAL and radio-loud quasars follow the same
alignment, with the same preferred polarization angle. However,
possible differences with other types of objects like BL Lac cannot be
excluded.  Also, it is important to remark that known polarization
differences between spectroscopically defined quasar types are not
washed out by the alignment effect. For example, the known difference
in polarization degree between BAL and non-BAL quasars is still valid
in region A1 as demonstrated in Paper~II.

Finally, it is interesting to recall that quasar radio polarizations
are usually not correlated to optical polarizations, and that radio
polarization vectors do not seem to show alignments as the optical
polarization vectors do (Paper~I, Vall\'ee \cite{VAL02}).

\section {Possible interpretations}
\label{sec:inter}

Possible interpretations of the alignment effect have been discussed
in Paper~I and II, and more recently by several authors (Jain et
al. \cite{JAI02,JAI04}, Bezerra et al. \cite{BEZ03}, Greyber
\cite{GRE03}, Ralston \& Jain \cite{RAS04}). They are further
discussed here in the light of the new results.

Since the alignments occur on extremely large scales and appear
connected on a sizeable fraction of the known Universe, one must seek
for global mechanisms acting at cosmological scales.  Possible
mechanisms must take into account the fact that the bulk of the
measured polarization is intrinsic to the quasars.  They may be
divided into two broad categories. First, the polarization angles may
be closely associated to the morphology of the objects, and the quasar
structural axes themselves are aligned on cosmological
scales. Alternatively, the polarization angles may be randomly
oriented at the source, and modified when the light propagates
throughout the Universe. Since both large-scale alignments and regular
rotation of the mean orientation must be explained, more than one
mechanism may contribute.

If we admit that quasar structural axes are coherently oriented at
such large scales, a global rotation of the Universe may be
invoked. It would transfer angular momentum to galaxies and quasars
during their formation, and, to some extent, correlate their
structural axes with the direction of the global rotation
(Li~\cite{LI98}).  In this case, one would expect the rotation axis to
be roughly perpendicular to the A1--A3 direction.  While it is not
excluded that complex effects like precession could be at the origin
of the redshift dependence of the mean orientation, a global rotation
would also induce a rotation of the polarization angles as a function
of the distance to the source (e.g. Obukhov~\cite{OBU00}).  The
mirror-like symmetry of the $\bar{\theta} - z$ relation illustrated in
Figs.~\ref{fig:tetaz2} and~\ref{fig:tetar2} would be accounted for by
a rotation axis close to the A1--A3 axis, which is different from the
direction needed to produce the alignments. An intermediate position
would then be required to explain both effects. From
Fig.~\ref{fig:tetar2}, we derive a universal angular velocity
$\omega_0 \gtrsim (\pi/2) \, H_0$ where $H_0$ is the Hubble constant,
in line with other estimates (K\"uhne \cite{KUH97},
Obukhov~\cite{OBU00}).  As a consequence of an inclined axis,
alignments should also be observed out of the A1--A3 regions.
Furthermore, the rotation of the polarization angle along the line of
sight would affect correlations between quasar polarization and
structural position angles, at least in some redshift ranges.
Interestingly enough, rotating cosmologies have been recently proposed
to explain possible anomalies in the CMB (Jaffe et al. \cite{JAF05}).
Another possible mechanism for aligning morphological axes could be
the effect of magnetic fields coherent over very large scales
(Reinhardt \cite{REI71}, Wasserman \cite{WAS78}, Battaner \& Lesch
\cite{BAT00}). Cosmological magnetic fields could make the expansion
of the Universe anisotropic (Berera et al. \cite{BEA04}) and then be
at the origin of a rotation of the polarization angles (Brans
\cite{BRA75}). If polarization vector alignments actually reflect
structural alignments, it is nevertheless difficult to explain the
alternance of coherently and randomly oriented polarization vectors
observed in Fig.~\ref{fig:statz2}.

The other possibility is that both the polarization vector alignments
and the rotation of the mean polarization angles are due to a
mechanism which affects the light on its travel towards the
observer. As shown in Paper~II, a small amount of polarization added
to randomly oriented polarization vectors can be at the origin of
coherent orientations of polarization angles\footnote{In fact, if we
had in mind to detect the effect of a small systematic polarization,
the study of polarization vector alignments is probably one of the
most sensitive methods. Indeed, since extragalactic objects are
usually intrinsically polarized at various levels
(Fig.~\ref{fig:hist_p}), the addition of a small systematic
polarization would be largely undetected, since it only slightly
broadens the distribution of the polarization degrees.  To some
extent, this is also true for low polarization objects
(e.g. radio-quiet quasars) because of the errors on the measurements
and the subsequent confusion with the polarization degree bias (errors
were not taken into account in the simulations of Paper~II). On the
other hand, a systematic polarization of a few tens of a percent
superimposed over randomly oriented polarization vectors do produce a
detectable effect in the distribution of the polarization angles.}
without scrambling too much the relation between polarization and
other quasar intrinsic properties.  Remarkably, a systematic
polarization and a rotation of the polarization angle are predicted by
photon--pseudoscalar mixing within a magnetic field, including a
quasi-periodic variation of the polarization along the line of sight
(e.g. Harari \& Sikivie \cite{HAR92}, Gnedin ~\cite{GNE94}, Das et
al.~\cite{DAS04}). Such an oscillation of the polarization added to
the quasar intrinsic polarization vectors would appear as a
quasi-periodicity in the alignment effect with redshift, in agreement
with the results of Sect.~\ref{sssec:regu}. Moreover, an associated
rotation of the polarization angles may be expected, as demonstrated
by recent simulations (Das et al. \cite{DAS04}).  Apparently,
photon--pseudoscalar mixing has the capability to explain most of the
characteristics of the alignment effect, with a coupling constant and
a magnetic field strength in agreement with current upper limits.  It
must be emphasized that this mechanism requires the existence of a
--hypothetical-- magnetic field organized on cosmological scales. The
symmetry of the $\bar{\theta} - z$ relation (Figs.~\ref{fig:tetaz2}
and~\ref{fig:tetar2}) would then correspond to the symmetry of the
magnetic field. Let us finally note that dust grains aligned in a
magnetic field can also produce some polarization, but would hardly
explain quasi-periodic alignments and a rotation of the mean
polarization angle.

Although still hypothetical, photon--pseudoscalar mixing within a
magnetic field appears as a promising interpretation, especially
because many of the observed characteristics of the alignment effect
were predicted, at least qualitatively.  Pseudoscalars may be related
to dark matter or dark energy, or be ejected by the quasars themselves
along with photons (Jain et al. \cite{JAI02}).  However, other
mechanisms like a global rotation of the Universe cannot be rejected,
and should be worked out in more details to see whether or not they
can reproduce the observations and constitute viable explanations.  In
addition to a better spatial sampling namely at higher redshifts, the
determination of a possible wavelength dependence of the polarization,
the behavior of circular polarization, the relation with quasar
morphological axes --especially along the A1--A3 axis-- would
definitely shed light on the responsible mechanism(s) and more
particularly on the photon--pseudoscalar mixing for which rather clear
predictions exist (Jain et al.~\cite{JAI02}, Das et al.~\cite{DAS04}).
Observations can also be readily performed to demonstrate the
existence of a possible preferred alignment axis, and its relation to
other tentative anisotropies in the Universe suggested either from the
CMB data or from other possible large-scale effects like the --still
controversial-- Birch effect (Birch~\cite{BIR82}, Jain \& Ralston
\cite{JAI99}).

\section {Conclusions}
\label{sec:conclu}
 
Based on new observations carried out during the period 2000 -- 2003,
we have constructed a new sample of quasar polarization measurements
in order to further investigate the extreme-scale alignments of quasar
polarization vectors discovered in Paper~I. The new sample contains
355 polarized quasars, i.e. more than two times the initial sample of
170 objects. Our goal was to firmly reassess the significance of the
alignment effect and to empirically derive constraints on possible
interpretations.

Using various, complementary, statistical methods, we demonstrate that
quasar polarization angles are definitely not randomly oriented over
the sky.  Polarization vectors appear coherently oriented over very
large spatial scales, in regions located at both low and high
redshifts and characterized by different preferred directions. These
properties make the alignment effect difficult to explain in terms of
local mechanisms, like a contamination by interstellar polarization in
our Galaxy.

Next, we tried to empirically characterize the effect and more
particularly its dependence on redshift. We found a regular alternance
of regions of coherently and randomly oriented polarization vectors
along the line of sight. We also showed that the mean polarization
angle is significantly correlated to redshift, rotating clockwise with
increasing redshift in North Galactic hemisphere and counter-clockwise
in the South Galactic one. Interestingly enough, the alignment effect
seems to be prominent along an axis not far from preferred directions
tentatively identified in Cosmic Microwave Background maps.

The fact that polarization vector alignments do occur on extremely
large scales and seem connected on a sizeable fraction of the known
Universe points towards a global mechanism acting at the scale of the
Universe. While several mechanisms like global rotation may, at least
partially, explain the alignment effect, we note that the observed
behavior remarkly corresponds to the dichroism and birefringence
predicted by photon-pseudoscalar oscillation within a magnetic field,
suggesting that we might have found a signature of either dark matter
or dark energy.

Such interpretations would have profound implications on our
understanding of the Universe and then certainly deserve further
studies. Fortunately, simple observations, although time consuming,
would readily allow to distinguish between possible interpretations,
the alignment effect then providing us with a new tool to probe the
Universe and its dark components.

%

\appendix
\section{Tables}
\label{sec:tables}
In Table~A.1 we give the polarization measurements for the full sample
of 355 quasars, i.e.  B1950 name/coordinates, the redshift $z$, the
polarization degree $p$ and its uncertainty $\sigma_p$, the
polarization position angle $\theta$ and its uncertainty
$\sigma_{\theta}$, and the references to the data.  References are
coded as follows: (0)~Hutsem\'ekers et al. \cite{HLR98}; (1)~Berriman
et al. \cite{BER90}; (2)~Stockman et al. \cite{STO84}; (3)~Moore \&
Stockman \cite{MOO84}; (4)~Impey \& Tapia \cite{IMP90}; (5)~Impey et
al. \cite{IMP91}; (6)~Wills et al. \cite{WIL92}; (7)~Visvanathan \&
Wills \cite{VIS98}; (8)~Schmidt \& Hines \cite{SCH99}; (9)~Lamy \&
Hutsem\'ekers \cite{LAM00}; (10)~Smith et al. \cite{SMI02}; (11)~Sluse
et al. \cite{SLU05}. References [1--6] were considered in Paper~I, and
[1--9] in Paper~II. The eight quasars from Impey \& Tapia
(\cite{IMP90}) with new redshift measurements and added to the final
sample are: B0118$-$272, B0138$-$097, B0301$-$243, B0426$-$380,
B1538$+$149, B1606$+$106, B1749$+$701, B2206$-$251. When better
polarization measurements are obtained, old values are replaced. Such
replacements were indicated in Paper~II. In addition, the polarization
measurements reported in Sluse et al. (\cite{SLU05}) for B1012$+$008,
B1048$-$090, B1216$-$010, B1216$+$069, B1222$+$228, B1545$+$210 and
B1617$+$175 supersede the values used in Paper~I and ~II.

\begin{table}\label{tab:sample}\caption{The sample of 355 polarized quasars}%
\begin{tabular}{lcrlrrr}\hline \\[-0.10in]%
Object  & $z$ & $p$ \ \ &  \ $\sigma_{p}$ & $\theta$ \ & $\sigma_{\theta}$ & Ref\\ 
(B1950) &  & (\%)  & (\%) & ($\degr$) & ($\degr$) & \\[0.05in]\hline \\[-0.10in]
 B0003$-$066 &   0.347 &    3.50 &   1.60 &   160 &   12 &     4 \\
 B0003$+$158 &   0.450 &    0.62 &   0.16 &   114 &    7 &     1 \\
 B0004$+$017 &   1.711 &    1.29 &   0.28 &   122 &    6 &     8 \\
 B0010$-$002 &   2.145 &    1.70 &   0.77 &   116 &   13 &     8 \\
 B0013$-$004 &   2.084 &    1.03 &   0.33 &   115 &   10 &     0 \\
 B0017$+$154 &   2.012 &    1.14 &   0.52 &   137 &   13 &     3 \\
 B0019$+$011 &   2.124 &    0.76 &   0.19 &    26 &    7 &     8 \\
 B0021$-$022 &   2.296 &    0.70 &   0.32 &   170 &   14 &     0 \\
 B0024$+$224 &   1.118 &    0.63 &   0.29 &    90 &   14 &     2 \\
 B0025$-$018 &   2.076 &    1.16 &   0.52 &   109 &   13 &     8 \\
 B0029$+$002 &   2.226 &    0.75 &   0.34 &   158 &   14 &     0 \\
 B0038$+$280 &   0.194 &    2.16 &   0.27 &   103 &    3 &    10 \\
 B0046$-$315 &   2.721 &   13.30 &   2.00 &   159 &    4 &     7 \\
 B0047$+$278 &   0.277 &    2.28 &   0.75 &    49 &    9 &    10 \\
 B0048$+$292 &   0.136 &    2.47 &   0.49 &    98 &    5 &    10 \\
 B0050$+$124 &   0.061 &    0.61 &   0.08 &     8 &    3 &     1 \\
 B0051$+$291 &   1.828 &    0.80 &   0.38 &   119 &   14 &     3 \\
 B0055$+$157 &   0.211 &    0.67 &   0.28 &    15 &   13 &    10 \\
 B0059$-$275 &   1.590 &    1.45 &   0.23 &   171 &    5 &     9 \\
 B0059$+$261 &   0.194 &    2.11 &   0.61 &   120 &    8 &    10 \\
 B0100$+$130 &   2.660 &    0.84 &   0.29 &   112 &   10 &     2 \\
 B0103$+$257 &   0.411 &    6.03 &   0.54 &   114 &    2 &    10 \\
 B0105$+$215 &   0.285 &    5.45 &   0.99 &   119 &    5 &    10 \\
 B0106$+$013 &   2.107 &    1.87 &   0.84 &   143 &   13 &     3 \\
 B0109$-$014 &   1.758 &    1.77 &   0.35 &    76 &    6 &     8 \\
 B0110$+$297 &   0.363 &    2.60 &   1.15 &    63 &   13 &     2 \\
 B0117$-$180 &   1.790 &    1.40 &   0.46 &    13 &   10 &     8 \\
 B0117$+$213 &   1.493 &    0.61 &   0.20 &   102 &    9 &     1 \\
 B0117$+$197 &   0.087 &    0.74 &   0.26 &   128 &   11 &    10 \\
 B0118$-$272 &   0.559 &   17.40 &   0.30 &   151 &    1 &     4 \\
 B0119$+$041 &   0.637 &    4.20 &   1.10 &    59 &    6 &     4 \\
 B0123$+$257 &   2.358 &    1.63 &   0.81 &   140 &   14 &     3 \\
 B0130$+$242 &   0.457 &    1.70 &   0.52 &   110 &    9 &     2 \\
 B0133$+$207 &   0.425 &    1.62 &   0.36 &    49 &    6 &     3 \\
 B0137$-$018 &   2.232 &    1.12 &   0.29 &    61 &    8 &     0 \\
 B0137$-$010 &   0.330 &    0.63 &   0.31 &   154 &   14 &     2 \\
 B0138$-$097 &   0.733 &    3.60 &   1.50 &   168 &   11 &     4 \\
 B0145$+$042 &   2.029 &    2.70 &   0.32 &   131 &    3 &     0 \\
 B0146$+$017 &   2.909 &    1.23 &   0.21 &   141 &    5 &     8 \\
 B0148$+$090 &   0.299 &    1.21 &   0.54 &   139 &   13 &     3 \\
 B0154$+$169 &   0.213 &    1.44 &   0.47 &    66 &    9 &    10 \\
 B0159$-$117 &   0.699 &    0.65 &   0.30 &     4 &   13 &     2 \\
 B0202$-$172 &   1.740 &    3.84 &   1.13 &    98 &    8 &     6 \\
 B0204$+$292 &   0.110 &    1.07 &   0.21 &   117 &    6 &    11 \\
 B0205$+$024 &   0.155 &    0.72 &   0.17 &    22 &    7 &     2 \\
 B0208$-$512 &   1.003 &   11.50 &   0.40 &    88 &    1 &     4 \\
 B0214$+$108 &   0.408 &    1.13 &   0.22 &   121 &    6 &     2 \\
 B0226$-$104 &   2.256 &    2.51 &   0.25 &   165 &    3 &     8 \\
 B0226$-$038 &   2.064 &    1.20 &   0.53 &    68 &   13 &     2 \\
 B0231$+$244 &   0.310 &    2.57 &   0.46 &    99 &    5 &    10 \\
 B0232$-$042 &   1.436 &    0.91 &   0.32 &   163 &   10 &     2 \\
 B0232$-$041 &   1.387 &    0.90 &   0.23 &    42 &    8 &    11 \\
 B0239$+$006 &   2.071 &    1.47 &   0.24 &   167 &    5 &    11 \\
 B0240$-$002 &   2.003 &    1.69 &   0.36 &    43 &    6 &    11 \\
 B0301$-$243 &   0.260 &   10.60 &   0.20 &    52 &    1 &     4 \\
\hline\end{tabular}\end{table}\addtocounter{table}{-1}
\begin{table}\caption{continued}\begin{tabular}{lcrlrrr}\hline \\[-0.10in] Object  & $z$ & $p$ \ \ & \ $\sigma_{p}$ & $\theta$ \ &  $\sigma_{\theta}$ & Ref\\ (B1950) &  & (\%)  & (\%) & ($\degr$)      & ($\degr$) & \\[0.05in]\hline \\[-0.10in]
 B0310$+$209 &   0.094 &    1.53 &   0.43 &   147 &    8 &    10 \\
 B0310$+$004 &   1.250 &    1.48 &   0.29 &   125 &    6 &    11 \\
 B0322$+$176 &   0.328 &    1.23 &   0.38 &   119 &    8 &    10 \\
 B0332$-$403 &   1.445 &   14.80 &   1.80 &   113 &    3 &     4 \\
 B0333$-$380 &   2.210 &    0.83 &   0.28 &    45 &   10 &     0 \\
 B0336$-$019 &   0.852 &   19.40 &   2.40 &    22 &    4 &     4 \\
 B0346$+$127 &   0.210 &    2.23 &   0.73 &    69 &    9 &    10 \\
 B0348$+$061 &   2.058 &    1.39 &   0.51 &   157 &   10 &     2 \\
 B0350$-$073 &   0.962 &    1.67 &   0.24 &    14 &    4 &     2 \\
 B0402$-$362 &   1.417 &    0.60 &   0.30 &    66 &   14 &     4 \\
 B0403$-$132 &   0.571 &    3.80 &   0.50 &   170 &    4 &     4 \\
 B0405$-$123 &   0.574 &    0.83 &   0.16 &   136 &    5 &     2 \\
 B0414$-$060 &   0.781 &    0.78 &   0.22 &   146 &    8 &     2 \\
 B0420$-$014 &   0.915 &   11.90 &   0.50 &   115 &    1 &     4 \\
 B0422$-$380 &   0.782 &    6.20 &   3.00 &   173 &   14 &     7 \\
 B0426$-$380 &   1.030 &    1.80 &   0.40 &    90 &    7 &     4 \\
 B0438$-$436 &   2.852 &    4.70 &   1.00 &    27 &    6 &     4 \\
 B0446$-$208 &   1.896 &    0.61 &   0.24 &   177 &   12 &    11 \\
 B0448$-$392 &   1.288 &    2.90 &   1.00 &    49 &   10 &     7 \\
 B0451$-$282 &   2.559 &    1.80 &   0.50 &    66 &    9 &     4 \\
 B0454$-$234 &   1.009 &   27.10 &   0.50 &     3 &    1 &     6 \\
 B0506$-$612 &   1.093 &    1.10 &   0.50 &    83 &   12 &     4 \\
 B0537$-$441 &   0.894 &   10.40 &   0.50 &   136 &    1 &     4 \\
 B0759$+$651 &   0.148 &    1.45 &   0.14 &   119 &    3 &     8 \\
 B0804$+$499 &   1.430 &    8.60 &   0.70 &   179 &    2 &     4 \\
 B0836$+$710 &   2.170 &    1.10 &   0.50 &   102 &   12 &     4 \\
 B0839$+$187 &   1.270 &    1.74 &   0.53 &   100 &    9 &     6 \\
 B0844$+$349 &   0.064 &    0.63 &   0.13 &    26 &    6 &     1 \\
 B0846$+$156 &   2.910 &    0.80 &   0.21 &   151 &    8 &     9 \\
 B0847$+$175 &   0.343 &    2.09 &   0.23 &   100 &    3 &    10 \\
 B0848$+$163 &   1.932 &    1.37 &   0.54 &    27 &   11 &     2 \\
 B0850$+$140 &   1.110 &    1.05 &   0.50 &   106 &   14 &     3 \\
 B0851$+$202 &   0.306 &   10.80 &   0.30 &   156 &    1 &     4 \\
 B0855$+$143 &   1.048 &    5.31 &   2.12 &    30 &   11 &     3 \\
 B0856$+$172 &   2.320 &    0.70 &   0.24 &     0 &   10 &     9 \\
 B0903$+$175 &   2.776 &    0.93 &   0.29 &    60 &    9 &     0 \\
 B0906$+$430 &   0.670 &    3.80 &   0.40 &    53 &    2 &     4 \\
 B0906$+$484 &   0.118 &    1.08 &   0.30 &   148 &    8 &     2 \\
 B0907$+$264 &   2.920 &    0.74 &   0.22 &   100 &    9 &    11 \\
 B0913$+$213 &   0.422 &    0.81 &   0.31 &     1 &   12 &    10 \\
 B0915$+$214 &   0.149 &    6.30 &   0.14 &   154 &    0 &    10 \\
 B0923$+$392 &   0.699 &    0.91 &   0.35 &   102 &   11 &     3 \\
 B0932$+$501 &   1.914 &    1.39 &   0.16 &   166 &    3 &     8 \\
 B0946$+$301 &   1.216 &    0.85 &   0.14 &   116 &    5 &     8 \\
 B0953$+$254 &   0.712 &    1.45 &   0.33 &   127 &    7 &     6 \\
 B0954$+$556 &   0.901 &    8.68 &   0.82 &     4 &    3 &     6 \\
 B0954$+$658 &   0.368 &   19.10 &   0.20 &   170 &    1 &     5 \\
 B0958$+$220 &   0.248 &    1.25 &   0.52 &   130 &   13 &    10 \\
 B1000$+$277 &   1.283 &    0.75 &   0.20 &    72 &    8 &    11 \\
 B1001$+$054 &   0.161 &    0.77 &   0.22 &    74 &    8 &     1 \\
 B1004$+$130 &   0.240 &    0.79 &   0.11 &    77 &    4 &     1 \\
 B1009$-$028 &   2.745 &    0.95 &   0.30 &   178 &    9 &     0 \\
 B1009$+$023 &   1.350 &    0.77 &   0.19 &   137 &    7 &     9 \\
 B1011$+$091 &   2.266 &    1.54 &   0.23 &   136 &    4 &     8 \\
 B1011$+$200 &   0.110 &    0.67 &   0.12 &    98 &    5 &    10 \\
\hline\end{tabular}\end{table}\addtocounter{table}{-1}
\begin{table}\caption{continued}\begin{tabular}{lcrlrrr}\hline \\[-0.10in] Object  & $z$ & $p$ \ \ & \ $\sigma_{p}$ & $\theta$ \ &  $\sigma_{\theta}$ & Ref\\ (B1950) &  & (\%)  & (\%) & ($\degr$)      & ($\degr$) & \\[0.05in]\hline \\[-0.10in]
 B1012$+$008 &   0.185 &    0.62 &   0.14 &   112 &    7 &    11 \\
 B1015$+$017 &   1.455 &    0.86 &   0.22 &   159 &    7 &    11 \\
 B1024$+$125 &   0.231 &    1.83 &   0.31 &   141 &    4 &    10 \\
 B1029$-$014 &   2.038 &    1.13 &   0.31 &   121 &    8 &     0 \\
 B1038$+$064 &   1.270 &    0.62 &   0.24 &   149 &   11 &     2 \\
 B1048$-$090 &   0.345 &    0.65 &   0.15 &   120 &    7 &    11 \\
 B1049$+$616 &   0.422 &    0.83 &   0.34 &   176 &   12 &     2 \\
 B1051$-$007 &   1.550 &    1.90 &   0.19 &    90 &    3 &     9 \\
 B1055$+$018 &   0.888 &    5.00 &   0.50 &   146 &    3 &     4 \\
 B1100$+$772 &   0.313 &    0.71 &   0.22 &    76 &    8 &     1 \\
 B1114$+$445 &   0.144 &    2.37 &   0.18 &    96 &    2 &     1 \\
 B1115$+$080 &   1.722 &    0.68 &   0.27 &    46 &   12 &     0 \\
 B1118$-$056 &   1.297 &    1.24 &   0.48 &    91 &   12 &    11 \\
 B1120$+$019 &   1.465 &    1.95 &   0.27 &     9 &    4 &     0 \\
 B1122$-$132 &   0.458 &    1.52 &   0.15 &   109 &    3 &    11 \\
 B1124$-$186 &   1.048 &   11.68 &   0.36 &    37 &    1 &    11 \\
 B1127$-$145 &   1.187 &    1.30 &   0.40 &    23 &   10 &     4 \\
 B1127$-$130 &   0.634 &    1.32 &   0.13 &    46 &    3 &    11 \\
 B1128$+$315 &   0.289 &    0.95 &   0.33 &   172 &   10 &     2 \\
 B1131$-$171 &   1.618 &    0.84 &   0.26 &    43 &    9 &    11 \\
 B1133$+$009 &   1.550 &    1.15 &   0.30 &    15 &    8 &    11 \\
 B1134$+$015 &   0.430 &    1.12 &   0.26 &   164 &    7 &    11 \\
 B1145$-$071 &   1.342 &    1.08 &   0.24 &    52 &    6 &    11 \\
 B1145$-$071 &   1.345 &    1.00 &   0.41 &   120 &   13 &    11 \\
 B1147$+$004 &   1.596 &    1.57 &   0.22 &   156 &    4 &    11 \\
 B1151$+$117 &   0.180 &    0.72 &   0.18 &   100 &    7 &     9 \\
 B1156$+$295 &   0.729 &    2.68 &   0.41 &   114 &    4 &     6 \\
 B1157$-$239 &   2.100 &    1.33 &   0.17 &    95 &    4 &     9 \\
 B1157$+$014 &   1.990 &    0.76 &   0.18 &    39 &    7 &     9 \\
 B1200$+$268 &   0.478 &    0.65 &   0.15 &   177 &    7 &    11 \\
 B1202$-$262 &   0.786 &    0.86 &   0.20 &    67 &    7 &    11 \\
 B1203$+$155 &   1.630 &    1.54 &   0.20 &    30 &    4 &     9 \\
 B1203$+$006 &   2.331 &    0.94 &   0.15 &   123 &    5 &    11 \\
 B1205$+$146 &   1.640 &    0.83 &   0.18 &   161 &    6 &     9 \\
 B1207$-$213 &   0.457 &    0.69 &   0.16 &    99 &    7 &    11 \\
 B1207$-$001 &   1.860 &    1.44 &   0.35 &   167 &    7 &    11 \\
 B1208$+$322 &   0.388 &    1.03 &   0.24 &    26 &    7 &     2 \\
 B1212$+$147 &   1.621 &    1.45 &   0.30 &    24 &    6 &     0 \\
 B1212$+$002 &   1.041 &    2.40 &   0.32 &   103 &    4 &    11 \\
 B1214$+$014 &   2.017 &    0.96 &   0.24 &    83 &    7 &    11 \\
 B1215$+$127 &   2.080 &    0.62 &   0.24 &    17 &   12 &     9 \\
 B1215$-$002 &   0.420 &   23.94 &   0.70 &    91 &    1 &    11 \\
 B1216$-$010 &   0.415 &   11.20 &   0.17 &   100 &    1 &    11 \\
 B1216$+$069 &   0.334 &    0.60 &   0.13 &    87 &    6 &    11 \\
 B1219$+$127 &   1.310 &    0.68 &   0.20 &   151 &    9 &     9 \\
 B1219$+$044 &   0.965 &    5.56 &   0.15 &   118 &    1 &    11 \\
 B1221$+$177 &   1.354 &    0.81 &   0.19 &    26 &    7 &    11 \\
 B1222$-$016 &   2.040 &    0.80 &   0.22 &   119 &    8 &     9 \\
 B1222$+$037 &   0.960 &    2.51 &   0.22 &    98 &    2 &    11 \\
 B1222$+$216 &   0.435 &    1.52 &   0.13 &   167 &    3 &    11 \\
 B1222$+$228 &   2.058 &    0.92 &   0.14 &   169 &    4 &    11 \\
 B1224$+$001 &   1.543 &    0.62 &   0.28 &   160 &   14 &    11 \\
 B1228$+$010 &   1.720 &    1.46 &   0.40 &    41 &    8 &    11 \\
 B1229$+$204 &   0.064 &    0.61 &   0.12 &   118 &    6 &     1 \\
 B1231$+$133 &   2.386 &    0.74 &   0.32 &   162 &   14 &     0 \\
\hline\end{tabular}\end{table}\addtocounter{table}{-1}
\begin{table}\caption{continued}\begin{tabular}{lcrlrrr}\hline \\[-0.10in] Object  & $z$ & $p$ \ \ & \ $\sigma_{p}$ & $\theta$ \ &  $\sigma_{\theta}$ & Ref\\ (B1950) &  & (\%)  & (\%) & ($\degr$)      & ($\degr$) & \\[0.05in]\hline \\[-0.10in]
 B1231$+$012 &   1.532 &    1.35 &   0.23 &     2 &    5 &    11 \\
 B1232$+$134 &   2.363 &    2.02 &   0.35 &    98 &    5 &     0 \\
 B1235$-$182 &   2.190 &    1.02 &   0.18 &   171 &    5 &     9 \\
 B1235$+$089 &   2.885 &    2.29 &   0.29 &    21 &    4 &     0 \\
 B1239$+$099 &   2.010 &    0.82 &   0.18 &   161 &    6 &     9 \\
 B1244$-$255 &   0.633 &    8.40 &   0.20 &   110 &    1 &     4 \\
 B1244$-$014 &   0.346 &    1.12 &   0.36 &    60 &   10 &    11 \\
 B1246$-$057 &   2.236 &    1.96 &   0.18 &   149 &    3 &     8 \\
 B1246$+$377 &   1.241 &    1.71 &   0.58 &   152 &   10 &     2 \\
 B1252$+$119 &   0.870 &    2.51 &   0.56 &   129 &    6 &     6 \\
 B1253$-$055 &   0.536 &    9.00 &   0.40 &    67 &    1 &     4 \\
 B1254$+$047 &   1.024 &    1.22 &   0.15 &   165 &    3 &     1 \\
 B1255$-$316 &   1.924 &    2.20 &   1.00 &   153 &   12 &     4 \\
 B1255$+$237 &   0.259 &    1.22 &   0.13 &   105 &    3 &    10 \\
 B1256$-$220 &   1.306 &    5.20 &   0.80 &   160 &    4 &     7 \\
 B1256$-$175 &   2.060 &    0.91 &   0.19 &    71 &    6 &     9 \\
 B1256$-$229 &   1.365 &   22.32 &   0.15 &   157 &    1 &    11 \\
 B1257$+$168 &   0.080 &    1.68 &   0.14 &    43 &    2 &    10 \\
 B1258$+$013 &   1.902 &    0.72 &   0.31 &    64 &   13 &    11 \\
 B1259$-$003 &   1.672 &    1.37 &   0.20 &    35 &    4 &    11 \\
 B1302$-$102 &   0.286 &    1.00 &   0.40 &    70 &   11 &     7 \\
 B1302$+$005 &   1.912 &    0.84 &   0.28 &    34 &   10 &    11 \\
 B1303$+$308 &   1.770 &    1.12 &   0.56 &   170 &   14 &     3 \\
 B1303$-$250 &   0.738 &    0.91 &   0.17 &   105 &    5 &    11 \\
 B1304$-$119 &   0.294 &    0.83 &   0.18 &    73 &    6 &    11 \\
 B1304$+$239 &   0.275 &    2.45 &   0.63 &    37 &    7 &    10 \\
 B1305$+$001 &   2.110 &    0.70 &   0.22 &   151 &    9 &     9 \\
 B1307$-$168 &   1.173 &    0.87 &   0.20 &    52 &    7 &    11 \\
 B1308$+$326 &   0.996 &   12.10 &   1.50 &    68 &    3 &     4 \\
 B1309$-$216 &   1.491 &   12.30 &   0.90 &   160 &    2 &     4 \\
 B1309$-$056 &   2.212 &    0.78 &   0.28 &   179 &   11 &     0 \\
 B1309$+$235 &   1.508 &    1.10 &   0.16 &   166 &    4 &    11 \\
 B1318$+$290 &   0.549 &    0.61 &   0.28 &    51 &   13 &     2 \\
 B1320$-$003 &   1.827 &    1.13 &   0.21 &    15 &    5 &    11 \\
 B1321$+$294 &   0.960 &    1.20 &   0.27 &   111 &    6 &     2 \\
 B1322$+$659 &   0.168 &    0.81 &   0.22 &    90 &    8 &     1 \\
 B1325$+$008 &   1.876 &    1.11 &   0.23 &    66 &    6 &    11 \\
 B1326$+$124 &   0.203 &    2.30 &   0.37 &    38 &    4 &    10 \\
 B1328$+$307 &   0.849 &    1.29 &   0.49 &    47 &   11 &     3 \\
 B1331$-$011 &   1.867 &    1.88 &   0.31 &    29 &    5 &     0 \\
 B1333$+$286 &   1.910 &    5.88 &   0.20 &   161 &    1 &     9 \\
 B1334$-$127 &   0.541 &   10.60 &   0.50 &     8 &    1 &     4 \\
 B1335$+$023 &   1.356 &    0.69 &   0.20 &    19 &    8 &    11 \\
 B1335$-$061 &   0.625 &    0.74 &   0.25 &    91 &   10 &    11 \\
 B1339$-$180 &   2.210 &    0.83 &   0.15 &    20 &    5 &    11 \\
 B1340$+$289 &   0.905 &    0.81 &   0.35 &    45 &   12 &     2 \\
 B1346$+$222 &   0.062 &    2.88 &   0.22 &   105 &    2 &    10 \\
 B1347$+$539 &   0.976 &    1.73 &   0.81 &   161 &   13 &     6 \\
 B1351$+$640 &   0.087 &    0.66 &   0.10 &    11 &    4 &     1 \\
 B1354$-$152 &   1.890 &    1.40 &   0.50 &    46 &   10 &     4 \\
 B1354$+$213 &   0.300 &    1.42 &   0.31 &    81 &    6 &     1 \\
 B1356$+$301 &   0.113 &    4.83 &   0.24 &    18 &    1 &    10 \\
 B1359$-$058 &   1.996 &    0.68 &   0.16 &   101 &    7 &    11 \\
 B1402$+$436 &   0.324 &    7.55 &   0.22 &    33 &    1 &     8 \\
 B1406$+$010 &   1.999 &    3.91 &   0.28 &    30 &    2 &    11 \\
\hline\end{tabular}\end{table}\addtocounter{table}{-1}
\begin{table}\caption{continued}\begin{tabular}{lcrlrrr}\hline \\[-0.10in] Object  & $z$ & $p$ \ \ & \ $\sigma_{p}$ & $\theta$ \ &  $\sigma_{\theta}$ & Ref\\ (B1950) &  & (\%)  & (\%) & ($\degr$)      & ($\degr$) & \\[0.05in]\hline \\[-0.10in]
 B1411$+$442 &   0.089 &    0.76 &   0.17 &    61 &    6 &     1 \\
 B1413$+$117 &   2.551 &    2.53 &   0.29 &    53 &    3 &     8 \\
 B1416$-$129 &   0.129 &    1.63 &   0.15 &    44 &    3 &     1 \\
 B1416$+$067 &   1.439 &    0.77 &   0.39 &   123 &   14 &     2 \\
 B1424$+$273 &   1.170 &    1.35 &   0.25 &    80 &    5 &    11 \\
 B1425$+$267 &   0.366 &    1.42 &   0.23 &    74 &    5 &     1 \\
 B1429$-$008 &   2.084 &    1.00 &   0.29 &     9 &    9 &     0 \\
 B1435$-$067 &   0.129 &    1.44 &   0.29 &    27 &    6 &     1 \\
 B1443$+$016 &   2.450 &    1.33 &   0.23 &   159 &    5 &     9 \\
 B1451$+$141 &   0.139 &    0.81 &   0.29 &    73 &   11 &    10 \\
 B1452$-$217 &   0.773 &   12.40 &   1.50 &    60 &    3 &     7 \\
 B1453$-$109 &   0.940 &    1.64 &   0.54 &    59 &    9 &     3 \\
 B1458$+$718 &   0.905 &    1.41 &   0.60 &   108 &   12 &     6 \\
 B1459$+$236 &   0.258 &    3.07 &   0.46 &   154 &    4 &    10 \\
 B1500$+$084 &   3.940 &    1.15 &   0.33 &   100 &    9 &     9 \\
 B1502$+$106 &   1.839 &    3.00 &   0.60 &   160 &    5 &     4 \\
 B1504$-$166 &   0.876 &    5.30 &   0.70 &    52 &    4 &     4 \\
 B1508$-$055 &   1.191 &    1.51 &   0.46 &    67 &    9 &     2 \\
 B1510$-$089 &   0.361 &    1.90 &   0.40 &    79 &    6 &     4 \\
 B1512$+$370 &   0.371 &    1.10 &   0.23 &   109 &    6 &     1 \\
 B1514$+$231 &   0.190 &    1.02 &   0.28 &   150 &    8 &    10 \\
 B1514$+$191 &   0.190 &    9.37 &   0.08 &   103 &    0 &    10 \\
 B1516$+$188 &   0.187 &    0.67 &   0.22 &   131 &   10 &    10 \\
 B1522$+$155 &   0.628 &    7.90 &   1.46 &    32 &    5 &     3 \\
 B1524$+$517 &   2.873 &    2.71 &   0.34 &    94 &    4 &     8 \\
 B1532$+$016 &   1.420 &    3.50 &   0.20 &   131 &    2 &     4 \\
 B1538$+$149 &   0.605 &   17.40 &   0.50 &   145 &    1 &     4 \\
 B1538$+$477 &   0.770 &    0.90 &   0.14 &    65 &    4 &     1 \\
 B1540$+$197 &   0.228 &    1.94 &   0.17 &    42 &    3 &    11 \\
 B1545$+$210 &   0.266 &    1.15 &   0.13 &    18 &    3 &    11 \\
 B1548$+$056 &   1.426 &    4.70 &   1.10 &    14 &    7 &     4 \\
 B1548$+$216 &   0.373 &    0.66 &   0.24 &    83 &   12 &    10 \\
 B1552$+$085 &   0.119 &    1.88 &   0.23 &    75 &    3 &     1 \\
 B1556$+$335 &   1.650 &    1.31 &   0.47 &    70 &   10 &     8 \\
 B1606$+$106 &   1.226 &    2.10 &   0.90 &   134 &   12 &     4 \\
 B1611$+$343 &   1.401 &    1.68 &   0.67 &   134 &   11 &     3 \\
 B1612$+$266 &   0.395 &    1.24 &   0.56 &    81 &   13 &     2 \\
 B1617$+$175 &   0.114 &    0.67 &   0.13 &    84 &    6 &    11 \\
 B1633$+$382 &   1.814 &    2.60 &   1.00 &    97 &   11 &     4 \\
 B1634$+$224 &   0.211 &    2.34 &   0.40 &   102 &    4 &    10 \\
 B1635$+$119 &   0.146 &    0.82 &   0.38 &   175 &   13 &     2 \\
 B1637$+$574 &   0.745 &    2.40 &   0.80 &   170 &    9 &     5 \\
 B1641$+$399 &   0.594 &    4.00 &   0.30 &   103 &    2 &     4 \\
 B1642$+$690 &   0.751 &   16.60 &   1.70 &     8 &    3 &     4 \\
 B1656$+$571 &   1.290 &    1.34 &   0.31 &    51 &    7 &     6 \\
 B1657$+$186 &   0.170 &    6.30 &   0.73 &   162 &    3 &    10 \\
 B1657$+$213 &   0.596 &   11.11 &   0.80 &   109 &    2 &    10 \\
 B1658$+$247 &   0.509 &    1.56 &   0.29 &    83 &    5 &    10 \\
 B1712$+$261 &   0.163 &    0.86 &   0.33 &    64 &   12 &    10 \\
 B1714$+$281 &   0.524 &    6.08 &   1.28 &     6 &    6 &    10 \\
 B1721$+$343 &   0.206 &    0.74 &   0.16 &   143 &    6 &     2 \\
 B1739$+$522 &   1.375 &    3.70 &   0.20 &   172 &    2 &     4 \\
 B1749$+$701 &   0.770 &   11.50 &   0.30 &   112 &    1 &     4 \\
 B2105$-$065 &   0.644 &    1.12 &   0.22 &   147 &    6 &    11 \\
 B2115$-$305 &   0.980 &    3.40 &   0.40 &    67 &    3 &     7 \\
\hline\end{tabular}\end{table}\addtocounter{table}{-1}
\begin{table}\caption{continued}\begin{tabular}{lcrlrrr}\hline \\[-0.10in] Object  & $z$ & $p$ \ \ & \ $\sigma_{p}$ & $\theta$ \ &  $\sigma_{\theta}$ & Ref\\ (B1950) &  & (\%)  & (\%) & ($\degr$)      & ($\degr$) & \\[0.05in]\hline \\[-0.10in]
 B2118$-$430 &   2.200 &    0.66 &   0.20 &   133 &    9 &     9 \\
 B2121$+$050 &   1.878 &   10.70 &   2.90 &    68 &    6 &     4 \\
 B2128$-$123 &   0.501 &    1.90 &   0.40 &    64 &    6 &     7 \\
 B2128$-$088 &   1.983 &    0.61 &   0.27 &   171 &   14 &    11 \\
 B2129$-$072 &   2.048 &    1.78 &   0.32 &    44 &    5 &    11 \\
 B2131$-$021 &   0.557 &   16.90 &   4.00 &    93 &    1 &     4 \\
 B2132$-$011 &   1.660 &    0.83 &   0.25 &   113 &    9 &    11 \\
 B2135$-$147 &   0.200 &    1.10 &   0.40 &   100 &   10 &     7 \\
 B2139$-$085 &   0.570 &    0.79 &   0.22 &   160 &    8 &    11 \\
 B2141$-$495 &   1.440 &    0.63 &   0.25 &   131 &   12 &    11 \\
 B2141$+$040 &   0.463 &    0.84 &   0.25 &   111 &    9 &    11 \\
 B2144$-$362 &   2.081 &    0.66 &   0.28 &    46 &   13 &    11 \\
 B2145$+$067 &   0.990 &    0.60 &   0.20 &   138 &   11 &     4 \\
 B2149$-$200 &   0.424 &    2.29 &   0.31 &    31 &    4 &    11 \\
 B2154$-$200 &   2.028 &    0.75 &   0.28 &   145 &   12 &     0 \\
 B2155$-$152 &   0.672 &   22.60 &   1.10 &     7 &    2 &     4 \\
 B2201$-$185 &   1.814 &    1.43 &   0.51 &     7 &   10 &     8 \\
 B2203$-$188 &   0.619 &    1.26 &   0.29 &    31 &    7 &    11 \\
 B2203$-$215 &   0.577 &    0.99 &   0.30 &    47 &    9 &    11 \\
 B2204$-$540 &   1.206 &    1.81 &   0.26 &   130 &    4 &    11 \\
 B2206$-$251 &   0.158 &   20.10 &   0.80 &   128 &    1 &     4 \\
 B2208$-$173 &   1.210 &    1.00 &   0.24 &   148 &    7 &    11 \\
 B2213$-$283 &   0.946 &    0.84 &   0.23 &    98 &    8 &    11 \\
 B2215$-$508 &   1.356 &    0.81 &   0.22 &   164 &    8 &    11 \\
 B2216$-$038 &   0.901 &    1.10 &   0.40 &   139 &   11 &     4 \\
 B2216$-$091 &   0.750 &    0.72 &   0.31 &     1 &   14 &    11 \\
 B2219$+$196 &   0.366 &    7.19 &   1.14 &   109 &    4 &    10 \\
 B2219$+$197 &   0.211 &    0.95 &   0.23 &   138 &    7 &    10 \\
 B2223$-$052 &   1.404 &   13.60 &   0.40 &   133 &    1 &     4 \\
 B2223$+$197 &   0.147 &    1.38 &   0.56 &    58 &   13 &    10 \\
 B2225$-$055 &   1.981 &    4.37 &   0.29 &   162 &    2 &     0 \\
 B2226$-$411 &   0.446 &    0.82 &   0.32 &    57 &   12 &    11 \\
 B2227$-$088 &   1.562 &    9.20 &   0.87 &   173 &    3 &     6 \\
 B2227$-$445 &   1.326 &    5.26 &   0.48 &    18 &    3 &    11 \\
 B2230$+$025 &   2.147 &    0.68 &   0.29 &   119 &   14 &     0 \\
 B2230$+$114 &   1.037 &    7.30 &   0.30 &   118 &    1 &     4 \\
 B2232$-$488 &   0.510 &    3.66 &   0.26 &    10 &    2 &    11 \\
 B2240$-$370 &   1.830 &    2.10 &   0.19 &    28 &    3 &     9 \\
 B2240$-$260 &   0.774 &   14.78 &   0.21 &   131 &    1 &    11 \\
 B2243$-$123 &   0.630 &    1.25 &   0.26 &   156 &    6 &     6 \\
 B2245$-$328 &   2.268 &    2.30 &   1.10 &    73 &   13 &     4 \\
 B2247$+$140 &   0.237 &    1.39 &   0.38 &    75 &    8 &     2 \\
 B2247$+$015 &   1.128 &    1.11 &   0.25 &    82 &    7 &    11 \\
 B2251$+$113 &   0.323 &    1.00 &   0.15 &    49 &    4 &     2 \\
 B2251$+$158 &   0.859 &    2.90 &   0.30 &   144 &    3 &     4 \\
 B2251$+$244 &   2.328 &    1.34 &   0.67 &   113 &   14 &     3 \\
 B2251$+$006 &   1.150 &    0.89 &   0.26 &   129 &    9 &    11 \\
 B2253$-$115 &   1.330 &    0.81 &   0.23 &   130 &    8 &    11 \\
 B2254$+$024 &   2.090 &    1.67 &   0.75 &     2 &   13 &     6 \\
 B2255$-$282 &   0.926 &    2.00 &   0.40 &   112 &    6 &     4 \\
 B2300$+$254 &   0.331 &    4.38 &   1.16 &   140 &    7 &    10 \\
 B2301$+$060 &   1.268 &    3.69 &   0.26 &   163 &    2 &    11 \\
 B2302$-$279 &   1.435 &    0.82 &   0.21 &     9 &    7 &    11 \\
 B2308$+$098 &   0.432 &    1.14 &   0.16 &   105 &    4 &     1 \\
 B2317$-$006 &   1.889 &    1.85 &   0.30 &   164 &    5 &    11 \\
\hline\end{tabular}\end{table}\addtocounter{table}{-1}
\begin{table}\caption{continued}\begin{tabular}{lcrlrrr}\hline \\[-0.10in] Object  & $z$ & $p$ \ \ & \ $\sigma_{p}$ & $\theta$ \ &  $\sigma_{\theta}$ & Ref\\ (B1950) &  & (\%)  & (\%) & ($\degr$)      & ($\degr$) & \\[0.05in]\hline \\[-0.10in]
 B2320$-$035 &   1.411 &    9.56 &   0.20 &    90 &    1 &    11 \\
 B2326$-$477 &   1.302 &    1.00 &   0.30 &   103 &    8 &     4 \\
 B2326$-$502 &   0.518 &    3.92 &   0.33 &   164 &    2 &    11 \\
 B2332$-$017 &   1.184 &    4.86 &   0.19 &    92 &    1 &    11 \\
 B2333$-$101 &   1.760 &    0.99 &   0.34 &   160 &   10 &    11 \\
 B2335$-$027 &   1.072 &    3.55 &   0.30 &   110 &    2 &    11 \\
 B2340$-$036 &   0.896 &    0.87 &   0.25 &   130 &    8 &     2 \\
 B2341$-$235 &   2.820 &    0.64 &   0.20 &   122 &    9 &     9 \\
 B2342$+$120 &   0.199 &    1.01 &   0.24 &   127 &    6 &    10 \\
 B2344$+$184 &   0.138 &    1.01 &   0.32 &    88 &   10 &    11 \\
 B2345$-$167 &   0.576 &    4.90 &   1.50 &    70 &    8 &     4 \\
 B2345$+$002 &   1.946 &    0.91 &   0.30 &   134 &   10 &    11 \\
 B2346$-$365 &   0.541 &    0.64 &   0.25 &    29 &   12 &    11 \\
 B2347$-$105 &   1.310 &    1.05 &   0.29 &   106 &    8 &    11 \\
 B2349$-$010 &   0.174 &    0.91 &   0.21 &   143 &    7 &     2 \\
 B2350$+$008 &   2.156 &    1.59 &   0.26 &    27 &    5 &    11 \\
 B2351$-$154 &   2.665 &    3.73 &   1.56 &    13 &   12 &     2 \\
 B2353$+$283 &   0.731 &    1.43 &   0.54 &    76 &   11 &     3 \\
 B2353$-$008 &   2.936 &    1.81 &   0.34 &    16 &    5 &    11 \\
 B2354$-$117 &   0.949 &    2.00 &   0.40 &   105 &    6 &     4 \\
 B2354$+$002 &   0.410 &    0.67 &   0.30 &    74 &   14 &    11 \\
 B2355$-$534 &   1.006 &    3.70 &   0.60 &   126 &    4 &     4 \\
 B2356$-$006 &   1.757 &    1.46 &   0.33 &   158 &    7 &    11 \\
 B2357$-$129 &   0.868 &    4.12 &   0.20 &   151 &    1 &    11 \\
 B2358$+$022 &   1.872 &    2.12 &   0.51 &    45 &    7 &     8 \\
\hline\\[-0.2cm]\end{tabular}

\footnotesize{References: (0)~Hutsem\'ekers et al. \cite{HLR98};
(1)~Berriman et al. \cite{BER90}; (2)~Stockman et al. \cite{STO84};
(3)~Moore \& Stockman \cite{MOO84}; (4)~Impey \& Tapia \cite{IMP90};
(5)~Impey et al. \cite{IMP91}; (6)~Wills et al. \cite{WIL92};
(7)~Visvanathan \& Wills \cite{VIS98}; (8)~Schmidt \& Hines
\cite{SCH99}; (9)~Lamy \& Hutsem\'ekers \cite{LAM00}; (10)~Smith et
al. \cite{SMI02}; (11)~Sluse et al. \cite{SLU05}}

\end{table}

\end{document}